# Institutional ownership and liquidity commonality: evidence from Australia

**Reza Bradrania**[a,*], **Robert Elliott**[a,b] **and Winston Wu**[c,a]

[a] UniSA Business, University of South Australia, Adelaide, Australia
[b] Haskayne School of Business, University of Calgary, Calgary, Canada
[c] The University of Sydney Business School, University of Sydney, Australia

**Abstract**

We study the liquidity commonality impact of local and foreign institutional investment in the Australian equity market in the cross-section and over time. We find that commonality in liquidity is higher for large stocks compared to small stocks in the cross-section of stocks, and the spread between the two has increased over the past two decades. We show that this divergence can be explained by foreign institutional ownership. This finding suggests that foreign institutional investment contributes to an increase in the exposure of large stocks to unexpected liquidity events in the local market. We find a positive association between foreign institutional ownership and commonality in liquidity across all stocks, particularly in large- and mid-cap stocks. Correlated trading by foreign institutions explains this association. However, local institutional ownership is positively related to the commonality in liquidity for large-cap stocks only.

* Corresponding author. Tel: +61 8 8302 0523. Fax: +61 8 8302 0992. Email: reza.bradrania@unisa.edu.au.
We thank two anonymous referees for the valuable comments and suggestions that have helped us to make significant improvements to this paper. We also thank the Securities Industry Research Centre of Asia-Pacific (SIRCA) and the Australian Securities Exchange (ASX) for providing the data used in this study. This research was supported by the Australian Research Council (ARC DP130103517).

# 1. Introduction

The co-movement, or so-called 'commonality', in liquidity among stocks is important for many investors and policymakers[1], and a growing body of work has considered identifying this commonality and its supply-side and demand-side sources. The supply-side source of commonality in liquidity is related to the funding constraints of the financial intermediaries who provide less liquidity during large market declines or highly volatile episodes of the market (Hameed, Kang, and Viswanathan, 2010; Coughenour and Saad, 2004; Brunnermeier and Pedersen, 2009). The demand-side sources refer to correlated trading activity (Chrodia, Roll, and Subrahmanyam, 2000; Hasbrouk and Seppi, 2001), the level of institutional ownership (Kamara, Lou, and Sadka, 2008; Koch, Ruenzi, and Starks, 2009; Karolyi Lee and Van Dijk, 2012; Moshirian, Qian, Wee and Zhang, 2017; Deng, Li and Li, 2018), investor sentiment, protection and environment (Huberman and Halka, 2011; Karolyi, Lee and Van Dijk, 2012; Moshirian, Qian, Wee and Zhang, 2017)[2].

Previous demand-side literature predicts that stocks held to a large extent by a group of investors that tend to trade in the same direction with similar timing will likely experience large trade imbalances at the same points in time, and consequently show high commonality in their liquidity (Kamara, Lou, and Sadka, 2008; Koch, Ruenzi, and Starks, 2009). Almost all of the evidence to date in support of this hypothesis that relates institutional ownership and commonality in liquidity focuses on the US markets. Indeed, little is known in other countries and even less about how it varies over time and across different types of investors.[3] The latter is important, as trading patterns of different investor groups in a market might be different.

[1] Chordia, Roll and Subrahmanyam (2000); Hasbrouk and Seppi (2001) and Huberman and Halka (2001), among others, show that the liquidity of a stock is time-varying and driven by a common component in liquidity across stocks. This common factor has implications for asset pricing as it has been shown to be a source of the systematic liquidity risk that affects asset prices (e.g., Pastor and Stambaugh, 2003; Acharya and Pedersen, 2005; Korajczyk and Sadka, 2008; and Lee, 2011).

[2] Moshirian, Qian, Wee and Zhang (2017) also show that the economic and financial environment factors affect commonality in liquidity. Since these factors have impacts on funding supply, they are close to supply-side determinants of liquidity commonality.

[3] To our knowledge, there are only four studies that look at the relationship between the level of institutional ownership and commonality in liquidity. Moshirian, Qian, Wee and Zhang (2017) and Deng, Li and Li (2018) provide a global evidence for the association of foreign ownership and commonality in liquidity. Koch, Ruenzi, and Starks (2009) and Kamara, Lou, and Sadka (2008) investigate the impact of mutual fund ownership, and institutional ownership, respectively, on liquidity commonality in the US market.

In this paper, we furnish a better understanding of the demand-side explanation of the commonality in liquidity by focusing on institutional ownership structure and investor types. Specifically, we study the evolution of liquidity commonality in the cross-section of stocks and examine the role of local as well as foreign institutional investors in liquidity commonality over time using a unique holding dataset in the Australian market.

The Australian market is an ideal setting to investigate this topic because none of these market participants are dominant holders of shares. Our holding dataset shows that on average about 30% (25%) of shares are owned by foreign (local) institutions. However, we hypothesize that foreign investors are the main contributor to the commonality in liquidity in Australia.

The intuition for our hypothesis is as follows. Previous research that links commonality in liquidity to the correlated trading behaviour of institutional investors shows that stocks owned by institutions with high turnover exhibit higher liquidity commonality (e.g., Koch, Ruenzi, and Starks, 2009). A dominant group of local institutional investors in Australia, such as superannuation funds, tend to follow buy-and-hold strategies whereas foreign institutions are likely to trade more often. For example, the average daily turnover of foreign institutions, based on our holding dataset, is about 0.37% compared to 0.31% for local institutions for the period of 2007-2014. During the same period, the average turnover of stocks in the ASX200 (the largest 200 stocks on the Australian Securities Exchange (ASX)), which are large stocks and most likely dominated by foreign investors, is about 0.46% compared to 0.23% for non-ASX200 stocks.

Further, the dividend imputation system in Australia may impact the trading behaviour of foreign and local investors differently. Many companies pay out most of their earnings as dividends, which results in a high, attractive dividend yield.[4] Since Australian investors are entitled to imputation credits they tend to hold the stocks and receive the dividends instead of selling them for capital gains. Foreign investors, however, are not entitled to the imputation credit and are more active in trading (Le, Yin and Zhao, 2019). Consequently, we hypothesize that stocks with high foreign ownership exhibit stronger commonality in liquidity compared to stocks owned mainly by local investors.

We test this hypothesis using a dataset from the Clearing House Electronic Subregister System (CHESS) of ASX that includes all share holdings in all stocks in the Australian market that are available for trading[5], and examine the impact of local and foreign institutional ownership on commonality in liquidity. More specifically, we study the evolution of liquidity commonality in Australia using a dataset of 2,215 stocks trading from 1990 through 2016, and investigate the relation between the commonality in liquidity and stocks' institutional ownership structure, both cross-sectionally and over time.

[4] The average dividend yield over our sample period is 3.47%

We use the Amihud (2002) price impact measure of illiquidity and use the market model of liquidity (Chordia, Roll, and Subrahmanyam, 2000) to estimate the sensitivity of an individual stock's liquidity to variation in market liquidity. Throughout this paper, we refer to this sensitivity as 'commonality in liquidity', 'liquidity beta' and 'systematic liquidity risk' interchangeably. We find that the commonality in liquidity is higher for large firms compared to the small firms in the cross-section, and the spread between the two has increased in recent times.

The divergence of commonality in liquidity between large and small stocks can be related to institutional basket or index trading, which involves large rather than small stocks (Koch, Ruenzi, and Starks, 2016; and Chordia, Roll, and Subrahmanyam, 2000). To investigate this explanation, we use CHESS data that includes the daily ownership of local and foreign institutions and show that commonality in liquidity increases with institutional ownership in the cross-section of stocks. The results are robust after controlling for different sets of control variables. We further segregate institutional investors into foreign and local institutions and find that liquidity betas increase with foreign institutional ownership across all stocks as well as large stocks, whereas for local institutions, the positive association is found only in large stocks. We also show that, in time series, the difference between the levels of foreign institutional ownership of large and small firms explains the spread in their liquidity commonality. There is a positive association between this spread and the growth of foreign institutional ownership, but not local institutional ownership.

Previous literature has documented correlated trading by institutional investors (Coval and Stafford, 2007; Greenwood and Thesmar, 2011; Anton and Polk, 2014; and Koch, Ruenzi, and Starks, 2016). Correlated trading can arise when stocks are owned by the same investors and are traded at the same time, or when investors face correlated liquidity innovations (e.g., Koch, Ruenzi, and Starks, 2016). It can also happen when institutions trade on the same information and in the same direction. For example, Sias (2004) documents herding behaviour among institutions in the US that results in correlated trading.

[5] Some 30% of shareholdings are held in issuer sponsored registers and are not directly accessible for trading. As soon as shares are transferred with the intention to be traded they are included in our data from CHESS.

We therefore examine whether foreign institutions contribute to commonality of liquidity through correlated trading. We follow Coughenour and Saad (2004) and Koch, Ruenzi, and Starks (2016) and define the sensitivity of changes in a firm's liquidity to movements in liquidity of a portfolio with high foreign ownership as a proxy for correlated trading. We show that stocks with higher foreign ownership experience more correlated trading both in the time-series and cross-sectional analysis. The correlated trading by foreign investors is more prevalent among large stocks compared to medium or small stocks. We further show that these results are robust to a variety of robustness tests. The findings suggest that the liquidity of stocks co-moves with the liquidity of a stock with high buying or selling pressures from foreign institutions, and that foreign institutional investors trade in a correlated way across all stocks.

Few research studies relate foreign and local institutional investors to commonality in liquidity. Karolyi, Lee and Van Dijk (2012), Moshirian, Qian, Wee and Zhang (2017) and Deng, Li and Li (2018) examine the sources of commonality in liquidity in an international setting. Karolyi, Lee and Van Dijk (2012) show that countries with higher foreign capital inflows in their equity market (which they associate with a greater presence of foreign institutional investors) experience higher commonality in liquidity. However, they define the capital inflows with respect to the US market, implying that this proxy can only be used for the US-based foreign institutional investors and not for foreign investors in non-US markets in their study (Deng, Li and Li, 2018). In the present study, we use investor holding data aggregated at the stock level as a direct measure of ownership, and provide evidence from outside the US for the role of foreign investors in driving liquidity commonality.

Deng, Li and Li (2018) find a significant negative relation between foreign institutional ownership and commonality in liquidity, whereas Moshirian, Qian, Wee and Zhang (2017) find no significant association between the two. Deng, Li and Li (2018) suggest corporate transparency as the mechanism through which foreign investors can reduce liquidity commonality around the world. Moshirian, Qian, Wee and Zhang (2017) and Deng, Li and Li (2018) use a global dataset of foreign and local institutional ownership, expressed as a percentage of capitalization of each market, from a similar data source. The fraction of each of these investor categories in their studies is very low[6]. Our concern is that their ownership datasets do not represent the typology of the investors in the markets included in their samples. We use the holding data directly from CHESS of ASX, and show that investment in the equity market by foreign institutions leads to a higher commonality in liquidity, and correlated trading is the mechanism through which foreign investors increase liquidity commonality.

Our finding of a positive relation between foreign ownership and commonality is not inconsistent with that of Deng, Li and Li (2018), since they show that there is a U-shaped association between foreign institutional ownership and liquidity commonality. They indicate that as foreign institutional ownership increases, its negative effect on stock liquidity commonality attenuates and becomes positive when foreign institutional ownership increases past a certain threshold. However, our study is distinct from theirs. We find correlated trading as the channel through which foreign ownership increases liquidity commonality, whereas they document corporate transparency as the mechanism by which foreign investors impact commonality. In an unreported analysis we find no evidence of corporate transparency as a channel through which foreign investors impact liquidity commonality in Australia.

This paper provides direct evidence on the impact of foreign and local ownership on liquidity commonality. We contribute to the literature of commonality in liquidity in different ways. First, prior studies on institutional ownership, as a demand-side source, or evolution of liquidity commonality, either concentrate in the US domestic market (Koch, Ruenzi, and Starks, 2016; Kamara, Lou, and Sadka, 2008) or examine the impact of institutional ownership on liquidity commonality in the international setting using a global data of mix of developed and developing countries (Moshirian, Qian, Wee and Zhang, 2017 and Deng, Li and Li, 2018). In this paper, we use a unique dataset that includes comprehensive daily holding records for all types of investor participants on an important developed market, outside the US, and document that foreign ownership plays a dominant role in driving liquidity in commonality compared to local investors in the cross-section and over time in a domestic market. To our best knowledge, Moshirian, Qian, Wee and Zhang (2017) and Deng, Li and Li (2018) are only two studies that examine the liquidity commonality impact of foreign ownership level, but from a global perspective. Our work advances knowledge of the impact of institutional trading on liquidity commonality in a domestic market.

[6] For example, Deng, Li and Li (2018) report that the average of foreign institutional ownership over their sample period for the US and Australian stock markets is 4.50% and 5.87%, respectively (see Table 1). In Moshirian, Qian, Wee and Zhang (2017), the average of foreign ownership in the US (Australia) stock market is 1% (4%), whereas the average of local institutional ownership in the US (Australia) is 19% (4%) (see Table 1). However, according to the US Federal Reserve Board and Goldman Sachs, the average of foreign and local institutional ownership since 1950 is 16% and 50%, respectively, and has been growing significantly since 2000 (see https://www.nasdaq.com/articles/four-causes-of-the-u.s.-stock-market-bubble-2020-09-10). Our holding dataset of the Australian equity market from the CHESS registry show that the average foreign and local institutional ownership in the Australian equity market is 31.43% and 24.62%, respectively.

Second, we address the lack of research on commonality in liquidity in the Australian equity market. Fabre and Frino (2004), as the only study focusing on this phenomenon in Australia, report weak results for the existence of commonality in Australia. They use quoted spread as the illiquidity measure in their study. They also use one year of high frequency data and a small number of stocks, and do not relate commonality to institutional ownership or correlated trading. Moreover, they do not attempt to examine time-series variation in commonality in liquidity. In this paper, we provide a robust evidence on the existence of commonality in liquidity using a long time-series dataset and document the growth of commonality over time concentrated on large stocks. We further confirm a demand side explanation for commonality in liquidity in Australia: correlated trading by foreign institutional investors.

Our findings have various implications. They are important for traders and portfolio managers as variations in illiquidity may affect mispricing opportunities and portfolio choice. Also, the cross-sectional divergence in the commonality in liquidity of large and small stocks in Australia suggests that the ability to diversify liquidity shocks by holding large (small) stocks has decreased (increased) in recent times. Large stocks are increasingly more sensitive to the systematic liquidity risk than small stocks, which suggests that the Australian equity market has become more vulnerable to liquidity shocks since the year 2001. Furthermore, we show that foreign institutions play an important role in this divergence and contribute to this vulnerability by investing more in the large and mid-cap stocks.

The paper is organized as follows. Section 2 describes the data and variable construction. Section 3 explains the temporal behaviour of commonality in liquidity. Section 4 investigates the role of institutional investors and particularly foreign investors in liquidity commonality using CHESS data. Section 5 examines correlated trading by foreign investors, and Section 6 concludes.

## 2. Data and variable construction

We investigate the evolution of commonality in liquidity over time in Australia and the role of investors in its time variation. We obtain daily trading data (close, spread, share volume, dollar volume, total number of shares) for all stocks listed on the ASX between January 1990 and December 2016 from the ACRD and End of Day databases. We collect monthly shares outstanding prior to 2000 from the SPPR database. All databases are provided by the Securities Industry Research Centre of Asia-Pacific (SIRCA).[7]

We also employ a comprehensive dataset provided by the ASX that contains daily closing ownership balances for different categories of investors for all common stocks listed on the ASX from January 2007 to August 2014. The ownership balances include all holdings available for trading.[8] The ASX constructs the dataset from the holding data in CHESS, which is the central register system that the ASX operates to register and settle share trading and to transfer share titles. Shareholders can also choose to register on an issuer-sponsored sub-register. However, these investors must first have their shares transferred to CHESS before they can trade them on the ASX. Holdings in 97.16% of the companies listed on the ASX are recorded in CHESS, covering about 72.78% of the total market capitalization of the Australian market. CHESS settles a transaction three banking days after the trading date (T+3) by simultaneously exchanging the legal ownership of shares and the payment between parties.[9] However, trading positions are reported to the clearing house if they are not reversed intraday (i.e., on a T+1 basis), so we effectively capture any positions held at least overnight.

The initial sample data from CHESS consists of the daily closing balance of holdings aggregated by investor category[10] for each of the 2,841 stocks. The data are aggregated into three investor categories, comprising local institutions (incorporated companies, banks, trusts, insurance companies, and superannuation funds), foreign institutions (nominees), and institutions (the combined holdings by foreign and local institutions). The nominee shareholdings recorded in CHESS are highly correlated to foreign institutional ownership reported by FactSet and Thomas-Reuters Global Ownership database (Le, Yin and Zhao 2019). If there are local interests behind some of the nominee holdings, they comprise a small fraction, as foreign holdings dominate the publicly known nominee ownership.[11] The CHESS records do not contain any categories of significant shareholders other than the three main groups we have identified.

[7] ASX End of Day only covers the period between January 2000 and December 2016. We collect share price and dollar volume data from the ACRD database and monthly shares outstanding from the SPPR database for the period between January 1990 and December 1999 to complete our sample.
[8] Some 30% of shareholdings are held in issuer-sponsored registers and are not directly accessible for trading. As soon as shares are transferred with the intention to be traded, they will be included in the data from CHESS. Therefore, the data used in our study effectively captures all trades carried out on the ASX.
[9] The ASX moved to T+2 settlement on 7 March, 2016.
[10] Registered categories in CHESS are individuals, incorporated companies, banks, trusts, insurances and superannuation funds and foreign corporations and their nominees.

We measure illiquidity using Amihud (2002) illiquidity measure, which equals the absolute value of a stock's daily return divided by the dollar volume and captures the price impact of trades. This measure can be calculated at a daily frequency, which allows us to study a longer time period where intraday data are not available. Furthermore, the Amihud's measure has been found to be correlated with alternative liquidity measures using intraday data (e.g., Korajczyk and Sadka 2008; Hasbrouck 2009). To avoid the nonstationary nature of Amihud's measure, we follow Kamara, Lou, and Sadka (2008) and Koch, Ruenzi, and Starks (2016) and use logarithmic changes in the daily measure of Amihud (2002) illiquidity measure as follows:

$$\Delta illiq_{i,d} = ln\left[\frac{illiq_{i,d}}{illiq_{i,d-1}}\right] = ln\left[\frac{|r_{i,d}| / |dvol_{i,d}|}{|r_{i,d-1}| / |dvol_{i,d-1}|}\right] \quad (1)$$

where $r_{i,d}$ is the return on stock *i* on day *d* and $dvol_{i,d}$ is the dollar volume of stock *i* on day *d*. We apply the following filters in computing the daily change in stock illiquidity. First, we delete daily observations from the sample if the stock price crossed a minimum tick level during the day in either direction.[12] Next, we delete stocks with a close price less than $0.01 from the sample. Then, following Amihud (2002), we eliminate firm-days of $\Delta illiq_{i,d}$ observations in the top and bottom 1% of the sample. Finally, in a given quarter we delete from the sample any stocks with less than 25 valid observations remaining after applying previous filters. The resulting sample consists of 44,003 firm-quarter observations, with an average of 407 firms per quarter.

[Insert Table 1 about here]

[11] According to Reserve Bank of Australia and based on the data sourced by Australian Bureau of Statistics, foreign ownership of Australian equities is about 35% of market capitalization of ASX between 2007 and 2010 (Black and Kirdwood, 2010). Our data indicate similar level of foreign institutional holdings during the same period.

[12] The filter, proposed by Fabre and Frino (2004), adjusts the filter of Chordia, Roll, and Subrahmanyam (2000) to capture the characteristics of the ASX. The alternative filter deletes a stock on the day its average price drops below $2 to avoid the contaminating influence of the minimum tick size. Our results do not change using the alternative filter. On the ASX, the minimum tick level of a stock depends on the market price of the stock. Price ranges and the minimum tick level (in parentheses) are: up to 10c (0.1c), 10c up to $2.00 (0.5c), $2.00 up to $99,999,999.00 (1c).

Table 1 reports the summary statistics of our sample. At the beginning of each quarter, we sort all the stocks into terciles based on the market capitalization at the end of the previous quarter and assign the stocks into three size groups: small (bottom tercile), mid-cap (middle tercile) or large (top tercile). We break down the summary statistics for different size groups and present the results in Panel A.

Panel A of Table 1 shows that the average capitalization for a large stock is $5,153 million. The number plunges to $252 million for an average mid-cap stock and $47 million for an average small stock. As we expect, the average daily price and the average daily dollar volume of a stock increase with its market capitalization. The average price and dollar volume are significantly smaller for small stocks ($0.63 and $140,420), compared with $8.83 and $16,125,070 for an average large stock.

Panel B of Table 1 shows the average ownership percentage, computed as the number of shares held by each institutional investor category divided by the total number of shares outstanding. Institutions are dominant owners of stocks in Australia. Foreign and local institutions combined hold, on average, 56.05% of sample stocks. On average, foreign institutions own more shares than local institutions. They hold 31.43% of the Australian share market, followed by local institutions (24.62%). The institutional ownership is higher in large stocks (62%) compared to small stocks (48.5%). Foreign institutions, on average, hold a higher portion of large shares (41.83%) than local institutions (20.17%). The opposite is true for small stocks: local institutions are the dominant owners, with 29.45% of small shares, whereas foreign institutions own 19.06% of these stocks.

Panel C of Table 1 reports the changes in shareholdings by each institutional investor category over the period when ownership data are available (2007 to 2014). The total shareholdings by all institutional investors increase by 11.77% from a daily average percentage holding of 49.83% in 2007 to 61.61% in 2014, with foreign institutions making up the majority of the gain (7.91%). The average daily percentage holdings by foreign institutions increase from 26.08% in 2007 to 33.99% in 2014, whereas the average daily percentage holdings by domestic institutions increase by 3.86% from 23.76% in 2007 to 27.62% in 2014. We also report the average daily dollar holdings in Panel C. The total value of shareholdings by all institutional investors increases by 27.7% from a daily average of $760.7 billion in 2007 to $971.6 billion in 2014. The average daily dollar holdings by foreign (domestic) institutions increase by 22.81% (37.79%) from $511.3 ($249.4) billion in 2007 to $627.9 ($343.7) billion in 2014.

## 3. Commonality in liquidity over time in Australia

In this section, we investigate commonality in liquidity over time and its sensitivity to institutional, foreign and local ownership in the Australian Stock Market. We start by examining the time series of changes in market illiquidity, $\Delta illiq_{mkt,d}$, defined as the daily average of $\Delta illiq_{i,d}$ (as in Eq. 1) for all stocks, weighted by the market capitalizations as of the previous trading day (Chordia, Roll, and Subrahmanyam, 2000). We plot the time series of its quarterly volatility in Fig. 1.

[Insert Fig. 1 about here]

Fig. 1 shows that the volatility of the change in market illiquidity increases over the sample period. It shows that volatile movements in $\Delta illiq_{mkt,d}$ can be traced to notable stock market events that occur in Australia during the year. For example, three of the most significant spikes in the volatility of market illiquidity occurred in 1992, 2011, and 2015. These correspond to the early 1990s recession, when the unemployment rate in Australia rose to 11%, the European debt crisis, and China's stock market crash and the subsequent economic slowdown in Australia, respectively.

To estimate the extent to which changes in the illiquidity of an individual stock co-move with changes in the market illiquidity (i.e., commonality) over time, we employ a market model of liquidity following Chordia, Roll, and Subrahmanyam (2000). For each firm quarter, we regress the firm's daily changes in the Amihud (2002) illiquidity measure on the daily changes in the value-weighted illiquidity measure of the market portfolio:

$$\Delta illiq_{i,d} = \alpha_i + \beta_{L,i}\, \Delta illiq_{mkt,d} + \sum_{j=1}^{J} \delta_j\, Z_{ji,d} + \epsilon_{i,d} \quad (2)$$

where the regression coefficient $\beta_{L,i}$, the market liquidity beta, measures the sensitivity of changes in a stock's liquidity to movements in market liquidity. For each regression, we remove the firm of interest from the market portfolio in estimating its market liquidity beta. We follow Chordia, Roll, and Subrahmanyam (2000) and control for the variables ($Z_{ji,d}$) that may correlate with the firm's liquidity. These include lead and lag changes in the market portfolio; lead, lag, and contemporaneous market returns; and contemporaneous squared stock return.

For each firm year, we compute the mean of the quarterly market liquidity betas. Table 2 presents the annual time series of the cross-sectional means of market liquidity beta of all stocks as well as large and small stocks.

[Insert Table 2 about here]

All means are positive and significant for all stocks, as well as large and small stocks, though they are stronger, economically and statistically, for large stocks compared to small ones. We also report the percentage of firms with a positive beta. This percentage varies, with a range of 54–68%, 61–79% and 43–59% for all, large and small stocks, respectively. The results show that there is a commonality in liquidity in the Australian market. This is different from the findings of Fabre and Frino (2004), who use quoted spread, as a transaction cost measure of liquidity, and report weak results for the existence of commonality in Australia; however, our finding is consistent with previous studies that report liquidity commonality in other developed markets (e.g. Chordia, Roll, and Subrahmanyam, 2000; Brockman and Chung, 2002; Choe and Yang 2010; and Domowitz, Hansch, and Wang, 2005).

Consistent with the finding of Chordia, Roll, and Subrahmanyam (2000) and Kamara, Lou, and Sadka (2008) that smaller stocks are less sensitive to market-wide shocks in liquidity in the US, we also find that the market liquidity beta of large firms, in general, is higher than that of small firms in Australia over our whole sample period of 1990–2016. However, this difference is less marked for the periods before the year 2001, compared to post-2001, where the null hypothesis that $H_0: \beta_{L,large} - \beta_{L,small} = 0$ is rejected for almost every year (exceptions are 2002 and 2013).

We plot the time series of quarterly market liquidity beta for large and small firms in Fig. 2(a) and the time series of the spread in market liquidity beta between large and small firms in Fig. 2(b). As Fig 2(a) shows, the evolution of liquidity beta over time is different for large and small firms. The sensitivity of large firms to market liquidity decreases (increases) until (after) year 2001. The trend is opposite for the small firms, as their liquidity beta increases until year 2001, but decreases after 2001. Fig. 2(b) shows that the spread between liquidity betas of large and small firms, which peaks in the second quarter of 1990 at 0.89, while subject to seasonal variations, declines to its lowest value, −0.08, in 2000. It then gradually increases to 0.903 in the second quarter of 2016. The trend of the difference in liquidity commonality between large and small firms during pre- and post-2001 is driven by both the decreasing (increasing) and increasing (decreasing) trend of liquidity betas of large (small) firms during these two periods. The results suggest that during the post-2001 period, in general, large (small) firms became more (less) sensitive to the changes in market liquidity in the Australian market. During the pre-2001 period, large firms became less sensitive to changes in market liquidity while the small firms became more sensitive to changes in market liquidity.

[Insert Fig. 2 about here]

Next, we examine if the time series of the market liquidity beta presented in Fig. 2 exhibit any time trends. We follow approaches used by Kamara, Lou, and Sadka (2008) and first conduct Dickey-Fuller (1981) unit-root tests. Specifically, for each time series, we run the following regression:

$$\beta_{L,t} = \alpha + \rho\beta_{L,t-1} + \gamma t + v_t \quad (3)$$

where $\alpha$ is a drift, *t* is a time trend and $\beta_{L,t-1}$ is the first lag of the average market liquidity beta at time *t*. Since Fig. 2 reveals that the time series of market liquidity beta pre-2001 are likely to be different from that post-2001, we also run the regressions for two sub-sample periods: 1990–2000 and 2001–2016. Panel A of Table 3 reports the results from the unit-root tests.

[Insert Table 3 about here]

The null hypothesis that there is a unit-root is rejected at conventional levels for large and small firms for the entire sample period, as well as the two sub-sample periods. In addition, we reject the null hypothesis that there is a stochastic time trend at conventional levels for the time series of differences in market liquidity beta between large and small firms for all sample periods.

We then test if there is a deterministic time trend in the time series of market liquidity betas by running the following regressions:

$$\beta_{L,t} = \alpha + \gamma t + v_t \quad (4)$$

where $\alpha$ is a constant and *t* is a time trend. Panel B of Table 3 presents the estimates of $\alpha$ and $\gamma$ for the full sample period 1990–2016, as well as for the two sub-sample periods: 1990–2000 and 2001–2016. During the period between 2001 and 2016, the time series of market liquidity beta for large firms displays a positive time trend, whereas the time series for small firms exhibits a negative time trend. As a result, the time series of differences in average market liquidity beta between large and small firms has a statistically significant positive time trend. In contrast, for the period between 1990 and 2000, the time series of average $\beta_L$ for large firms displays a statistically significant negative time trend, while the time series of average $\beta_L$ for small firms has a positive time trend. Again, the results suggest that towards the year 2001, large firms became less sensitive to changes in market liquidity, while the opposite was true for small firms, resulting in a statistically significant downward time trend in the spread of market liquidity betas between large and small firms.

The increasing (decreasing) trend in systematic liquidity risk for large (small) firms in Australia since 2001 is consistent with findings in the US (Kamara, Lou, and Sadka, 2008; Gompers and Metrick, 2001; and Harford and Kaul, 2005) and can be explained by the growth of institutional trading over time. Chordia, Roll, and Subrahmanyam (2000) show that commonality in liquidity is driven by correlated trading of stocks by institutions with similar investment styles. The summary statistics results in Table 1, Panel B, show that in Australia, institutions on average hold more large shares than small shares and that foreign institutions are the dominant owners of the large stocks. Also, the results in Table 1, panel C show that foreign institutional ownership has increased more than local institutional ownership over time in Australia, suggesting that foreign investors might be a main contributor to commonality in liquidity. To test this, we further examine the trading activity of both local and foreign investors. We anticipate that local institutions in Australia are likely to be buy-and-hold investors, whereas foreign institutions are likely to trade more often. If this is true, we hypothesize that foreign institutions play a more important role in commonality in liquidity in Australia compared with the local institutions. The intuition for our hypothesis is that local institutional investors are dominated by superannuation funds who tend to be long-term investors with a buy-and-hold focus. Moreover, local investors are entitled to imputation credits and tend to hold the shares and receive the dividends instead of selling them for capital gains (Le, Yin and Zhao, 2019).

[Insert Table 4 about here]

To test the hypothesis that foreign investors trade more often than local institutional investors, we sort stocks into quintiles based on the fraction of ownership by each of the institutional investor categories at the end of the previous quarter, and present average daily turnover for each ownership quintile, as well as the mean difference between turnover of the top and the bottom quintiles and their respective *t*-statistics in Table 4, Panel A. The results show that the trading activity increases by foreign ownership monotonically, whereas this is not the case for the local investors. The difference between the turnover of the top and the bottom quintiles for both investor groups is significant; however, it is seven times greater for the foreign investors (0.279%, *t*-statistic=22.3) compared to local institutions (0.045%, *t*-statistic=5.6). We also present ownership-weighted turnover and holding periods for both investor groups in Panel B. For each quarter, we compute the cross-sectional average of turnover, weighted by ownership percentage, at the end of the previous quarter for both investor groups, and compute the cross-sectional holding period (in days) as the inverse of cross-sectional average of turnover. We report time series means of cross-sectional averages of turnover and holding period for each investor type in Panel B. The average daily turnover of holdings of foreign investors is 0.365%, which is greater than that of the local investors (0.309%). Moreover, foreign investors hold on to their shares for 53 days less than local investors, on average.

As an alternative approach to examine the implied trading activity of foreign investors, we focus on the turnover of the ASX200 stocks that are large and owned more by foreign investors than the non-ASX200 stocks. Panel C in Table 4 shows that the daily average turnover of the ASX200 (0.46%) is almost two times of non-ASX200 stocks (0.23%) during the period for which we have the holding data. Furthermore, the average holding period for ASX200 stocks is 221 days, which is about half the holding period for non-ASX200 stocks (451 days).

The results in Table 4 provide further support for our hypothesis that foreign investors trade more often than local institutions and, consequently, we postulate that foreign institutions are the main contributor to the commonality in liquidity compared to local investors in Australia. In the next section, we formally test this by examining the association between growth of institutional ownership in general, and foreign and local ownership in particular, and commonality in liquidity. However, our analysis is limited to the period 2007–2014, for which we have the ownership data. Unfortunately, we are not able to analyse the opposite time trend for commonality in liquidity before the year 2001 in Australia because we do not have ownership data for that period. However, one potential explanation for this, based on our hypothesis that foreign investors drive commonality, is related to the change in the participation of foreign institutions due to economic conditions in Australia during that period. The recession of the early 1990s had a lasting effect on the economy of Australia. For example, the unemployment rate reached 10.86% and it was not until 2000 that the unemployment rate returned to the pre-crisis level. As a result, foreign holdings of Australian shares dropped from a high of 61% in 1990 to only 35% in 2000, and subsequently rose to a 20-year high of 50% in 2012 (Australian Bureau of Statistics, 2019)[13]. In the next section, we formally examine the hypothesis that trading by institutional investors, especially foreign investors, contributes to commonality in liquidity in Australia.

# 4. Commonality in liquidity and foreign ownership

In this section, we investigate the relation between the sensitivity of changes in a stock's liquidity to movements in market liquidity (market liquidity beta) and the stock's share ownership structure. Since our ownership dataset starts in 2007 and ends in 2014, we are not able to examine if changes in ownership structure correlate with the time trend characteristics exhibited in the time series of market liquidity beta for the whole sample period of 1990–2016. On the other hand, our ownership sample period includes two major financial crises (the Global Financial Crisis and the European Debt Crisis), which resulted in both turbulent market conditions and substantial ownership structure changes. This allows us to make meaningful inferences using the available sample period. Furthermore, we avoid the problem of spurious correlations as the time series of market liquidity beta exhibits no time-trend during the period between 2007 and 2014.[14]

We focus our analysis on the institutional investors and further decompose institutional investors into two types: foreign (nominee) institutional investors and local institutional investors. To test the relation between market liquidity beta and share ownership structure in cross-section, we use the Fama and Macbeth (1973) approach and estimate the following regression in each quarter for each investor type:

$$\beta_{L,i,t} = \alpha + b_1 Own_{i,t-1} + b_2 Size_{i,t-1} + b_3 Illiq_{i,t-1} + \upsilon_{i,t} \quad (5)$$

where $\beta_{L,i,t}$ is the liquidity beta for firm $i$ in the current quarter, estimated as in Eq. 2. $Own_{i,t-1}$ and $Size_{i,t-1}$ are the ownership percentage (computed as the number of shares held by each institutional investor category divided by the total number of shares outstanding), and the natural logarithm of the firm's market capitalization (in millions), respectively, and both are calculated at the end of the previous quarter. $Illiq_{i,t-1}$ is the average daily Amihud (2002) illiquidity measure and is computed over the previous quarter. We include size and illiquidity as control variables in the regression as there is a high correlation between size and liquidity of the firms and institutional ownership (e.g., Chordia, Roll, and Subrahmanyam, 2000; Del Guercio, 1996; and Falkenstein, 1996).

[13] www.abs.gov.au
[14] Time trend test results during this sample period are available upon request.

[Insert Table 5 about here]

Table 5 presents the time-series averages of coefficients for different investor types across all stocks as well as for each size group. The *t*-statistics are adjusted for heteroskedasticity and autocorrelation with Newey and West (1987) standard errors with two lags. Model 1 shows the results for the regression of $\beta_L$ on institutional ownership only. The results show that stock ownership by institutional investors is associated with higher liquidity beta for all stocks as well as for large stocks. The slope for the ownership is 0.315 (*t*-statistic=9.44) for all stocks and 0.256 (*t*-statistic=7.39) for large stocks. The value of the coefficient of ownership is positive but marginally significant (insignificant) for mid-cap (small) stocks. These results hold after controlling for size and illiquidity in Model 2.

In Models 3–5, we test if stock ownership by different institutional investors, i.e., foreign and local institutions, has different impacts on liquidity commonality. Model 3 shows that higher stock ownership by foreign investors at the end of the previous quarter is associated with higher market liquidity beta in the current quarter for all stocks as well as for all size groups except small stocks, for which the results are insignificant. The slope of the ownership for all stocks is 0.11 (*t*-statistic=3.89), and 0.141 (*t*-statistic=2.93) and 0.102 (*t*-statistic=2.06) for large and mid-cap stocks, respectively. In Model 4, we investigate if local ownership is related to liquidity beta. The results indicate that higher market liquidity beta is associated with a higher fraction of local ownership for large stocks only. The slope of the local institutional ownership is 0.163 (*t*-statistic=2.32) for large stocks, but it is insignificant for all stocks as well as for other size groups. In Model 5, we include both local and foreign institutional ownerships in the regression and observe very similar results compared to when we include them separately in the regression: for all stocks, there is a positive and significant (insignificant) association between foreign (local) ownership and liquidity beta. For the large size group, the slopes for both local and foreign ownership are positive and significant, although the association is stronger between foreign ownership and liquidity beta. Finally, among mid-cap stocks there is a positive and (marginally) significant relation between foreign ownership and liquidity beta, but it is insignificant for local ownership.

Our results in Table 5 are consistent with the findings of Kamara, Lou, and Sadka (2008) and Koch, Ruenzi, and Starks (2016), who show that institutional investing is a source of commonality in stock liquidity in the US, and that this association is more prevalent among large stocks. In addition to these findings, we confirm our hypothesis that foreign investors contribute more to the commonality in liquidity in Australia. We show that greater investment in the stock market by foreign institutional investors leads to a higher market liquidity beta, whereas investment by local investors has a positive impact on the market liquidity beta only for large stocks. Our findings on the role of foreign institutions in liquidity commonality in Australia are consistent with the findings of Karolyi, Lee and Van Dijk (2012), who provide evidence that commonality in liquidity is higher in markets with higher foreign participation using a global dataset. However, they are in contrast with those of Deng, Li and Li (2018) and Moshirian, Qian, Wee and Zhang (2017), who also examine international markets. Deng, Li and Li (2018) find a significant negative relation between foreign institutional ownership and commonality in liquidity, while Moshirian, Qian, Wee and Zhang (2017) find no significant association between the two. Our finding of a positive relation between foreign ownership and commonality is not inconsistent with that of Deng, Li and Li (2018). They show that there is a U-shaped association between foreign institutional ownership and liquidity commonality and suggest that as foreign institutional ownership increases, its negative effect on stock liquidity commonality attenuates. It becomes positive when foreign institutional ownership increases past a certain threshold.

We then examine if the cross-sectional difference between the market liquidity betas of large and small firms over time is associated with growth in institutional ownership.[15] We regress the time series of the spread of the liquidity betas between large and small firms on the difference in institutional ownerships in those size groups:

[15] Note that our ownership data is limited to 2007–2014, but covers about half of the 2001–2016 period for which the increasing spread in liquidity beta are shown in Figure 2.

$$\beta_{L,large,t} - \beta_{L,small,t} = \alpha + b_1 t + b_2(Own_{large,t-1} - Own_{small,t-1})$$
$$+ b_3(Size_{large,t-1} - Size_{small,t-1})$$
$$+ b_4(Illiq_{large,t-1} - Illiq_{small,t-1}) + \upsilon_{i,t} \qquad (6)$$

where $\beta_{L,large,t}$ and $\beta_{L,small,t}$ are the averages of quarterly market liquidity beta in the current quarter for large and small firms, respectively. $Own_{large,t-1}$ and $Own_{small,t-1}$ are the averages of the number of shares held by the investor type (institutions, foreign institutions or local institutions) divided by the number of shares outstanding for large and small firms, respectively. $Size_{large,t-1}$ and $Size_{small,t-1}$ are the averages of the natural logarithm of the firm's market capitalization (in millions) for large and small firms, respectively. $Illiq_{large,t-1}$ and $Illiq_{small,t-1}$ are the averages of daily Amihud (2002) illiquidity measure for large and small firms, respectively, estimated over the previous quarter.

[Insert Table 6 about here]

Table 6 presents the results for institutions, which are then further decomposed into foreign and local institutions. Model 1 shows the results for the regression of the spread between the liquidity betas of large and small firms on the difference in institutional ownership. The coefficient is positive and significant and becomes stronger (1.945 and *t*-statistic=3.16) after we control for size and illiquidity spreads in Model 2. We further focus on the impact of foreign and local institutional ownership on the spread of the liquidity beta in Models 3 to 7. Models 3 and 4 show a significant relation between foreign ownership and the commonality spread, with a slope of 1.571 (*t*-statistic=2.96) in Model 4, whereas Models 5 and 6 show this relation is insignificant for local ownership, before and after taking controls into account. The significant (insignificant) results for foreign (local) institutions are robust when we include both foreign and local ownerships as well as controls in Model 7. Our results in Table 6 suggest that the cross-sectional spread in market liquidity betas between large and small firms is positively associated with differences in foreign institutional ownership between large and small stocks.

In summary, in this section we show that in the cross-section, the fraction of foreign institutional ownership of a firm is positively associated with its market liquidity beta. Higher local institutional ownership leads to higher market liquidity beta, but only for large stocks. In the time series, we find that liquidity commonality increases following the growth in foreign investing, but there is an insignificant association between liquidity commonality and local institutional ownership.

Previous literature (Kamara, Lou, and Sadka 2008; and Koch, Ruenzi, and Starks 2016) suggests that correlated trading by institutional investors contributes to commonality of liquidity. If a group of investors holds a set of stocks, they tend to trade in the same direction at the same time due to common ownership or liquidity shocks. As a result, these stocks are likely to experience large trade imbalances, and be characterized by strong comovements in their liquidity. In the next section, we examine if correlated trading explains the commonality in liquidity among stocks with higher institutional ownership, particularly by foreign institutions, in Australia.

# 5. Foreign institutions, correlated trading and commonality in liquidity

Chordia, Roll, and Subrahmanyam (2000) suggest that commonality in liquidity can be traced to common variation in trading activity. Following Koch, Ruenzi, and Starks (2016), we assume that stocks with high ownership by investor type are traded more by the same investor category.[16] Consequently, we use foreign institutional ownership aggregated at the stock level as a proxy for the foreign investor trading. To test the hypothesis that correlated trading explains commonality in liquidity among the stocks with higher institutional ownership we employ the two-step approach suggested by Coughenour and Saad (2004) and Koch, Ruenzi, and Starks (2016). First, we measure a stock's liquidity comovement with the liquidity of a portfolio with high foreign ownership percentage, and then estimate the relationship of that comovement to the stock's ownership by foreign investors.

Specifically, each quarter, we sort all stocks into deciles based on the foreign ownership percentage in the previous quarter. The ownership percentage is computed as the number of shares held by foreign investors divided by the total number of shares outstanding. The top decile is the portfolio of stocks with high foreign ownership. We calculate the value-weighted Amihud (2002) illiquidity measure of this portfolio ($illiq_{HI,d}$). Now, for each firm quarter, we regress the firm's daily changes in Amihud (2002) illiquidity measure ($\Delta illiq_{i,d}$) on the daily changes in value-weighted illiquidity measure of the portfolio with high foreign ownership ($\Delta illiq_{HI,d}$), as well as daily changes in market illiquidity and control variables.

[16] As we explained in section 3, our results in Table 1, panels B and C, and Table 4 are consistent with this assumption.

$$\Delta illiq_{i,d} = \alpha_i + \beta_{HI,i}\, \Delta illiq_{HI,d} + \beta_{L,i}\, \Delta illiq_{mkt,d} + \sum_{j=1}^{J} \delta_j\, Z_{ji,d} + \epsilon_{i,d} \quad (7)$$

where the regression coefficient $\beta_{HI,i}$ measures the sensitivity of changes in a firm's liquidity to movements in liquidity of a portfolio with high foreign ownership. We henceforth refer to $\beta_{HI,i}$ as high ownership liquidity beta or simply $\beta_{HI}$. The coefficient $\beta_{L,i}$ is the market liquidity beta. We control for the market liquidity beta to focus only on commonality of stocks with high ownership proxied for high trading. $Z_{ji,d}$ is a set of controls that may correlate with the firm's liquidity. They represent lead and lag changes in the market portfolio as well as the portfolio with high foreign ownership; lead, lag, and contemporaneous market returns; and contemporaneous squared stock return.

[Insert Fig. 3 about here]

Fig. 3 plots the time series of the cross-sectional average of high foreign ownership liquidity beta, as well as the percentage of positive high foreign ownership liquidity betas. The average $\beta_{HI}$ is positive in most quarters. Similarly, the percentage of the positive betas in the cross-section exceeds 50% in most quarters for all institutional investor types over time. On average, a positive $\beta_{HI}$ suggests that the liquidity of an individual stock is positively associated with the liquidity of a portfolio with high foreign ownership.

Now, we test if the $\beta_{HI}$ covaries positively with the stock ownership by foreign investors. In Section 4, we found that across all stocks as well as large and mid-size stocks, the fraction of foreign institutional ownership of a stock co-moves positively with its market liquidity beta ($\beta_L$). Consequently, if these comovements are due to the correlated trading by these types of investors, we expect that the foreign institutional ownership would be positively associated with $\beta_{HI}$ of a portfolio with high ownership by foreign institutions.

To examine this, we sort stocks into quintiles based on the fraction of foreign ownership at the end of the previous quarter. Table 7 presents equally-weighted average $\beta_{HI}$ for foreign ownership quintile, as well as the mean difference between $\beta_{HI}$ of the top and the bottom quintiles and their respective *t*-statistics. Panel A reports the results across all stocks, whereas Panel B includes the double sort results, first based on size and then ownership.

[Insert Table 7 about here]

The results in Table 7, Panel A, show that $\beta_{HI}$ increases monotonically with foreign ownership. The portfolio with the highest foreign institutional ownership has an average $\beta_{HI}$ of 0.142, compared with 0.025 for the portfolio with the lowest foreign ownership. The $\beta_{HI}$ spread between high and low foreign ownership is 0.117 (*t*-statistic = 5.37).

We also report dependent double-sort results based on size and foreign ownership in Table 7 Panel B. The results show that for large stocks, the liquidity of stocks with high foreign institutional ownership comoves more heavily with the liquidity of the high foreign ownership portfolio than stocks with low foreign institutional ownership. There is no significant difference in $\beta_{HI}$ between the highest and lowest quintiles sorted by foreign institutional ownership among small and mid-size stocks. In summary, the univariate and double sort results in Table 7 show that the foreign institutional ownership is positively associated with $\beta_{HI}$ of a portfolio with high ownership by foreign institutions. These results indicate that this association is due to the correlated trading by foreign investors.

We further follow the Fama and Macbeth (1973) approach and estimate the following regression across all stocks in the cross-section:

$$\beta_{HI,i,t} = \alpha + b_1 Fown_{i,t-1} + b_2 Size_{i,t-1} + b_3 Illiq_{i,t-1} + \upsilon_{i,t} \quad (8)$$

where $\beta_{HI,i,t}$ is estimated in Eq. 7, and $Fown_{i,t-1}$is the foreign ownership and computed as the number of shares held by foreign investors divided by the total shares outstanding. $Size_{i,t-1}$is the natural logarithm of the firm's market capitalization (in millions). Size and foreign ownership are computed at the end of the previous quarter. $Illiq_{i,t-1}$ is the average daily Amihud (2002) illiquidity measure, computed over the previous quarter. Table 8 presents time-series averages of coefficients for all stocks as well as for each size group. The *t*-statistics are adjusted using the Newey and West (1987) method with two lags.

[Insert Table 8 about here]

The regression results in Table 8 show that foreign ownership has a positive and statistically significant coefficient across all stocks (Models 1 and 2) as well as large size group (Models 3 and 4). The findings in this section are consistent with the explanation that foreign institutions tend to trade in a correlated fashion across all stocks and particularly large stocks and contribute to commonality in liquidity. Deng, Li and Li (2018) suggest that corporate transparency is the mechanism through which foreign investors can impact commonality around the world. In an unreported analysis[17], we find no evidence of corporate transparency as a channel through which foreign investors drive liquidity commonality in Australia.

As an alternative and robustness test to examine correlated trading among foreign investors, we focus on ASX200 constituents that are likely dominated by foreigners in terms of trading and that probably show abundant correlated trading versus non-ASX200 stocks. We compute and compare statistics on various proxies for correlated trading for ASX-200 as well non-ASX200 stocks. The proxies for correlated trading include dollar trading volume, turnover, daily return autocorrelation, and a trading volume comovement measure. The first two proxies capture the trading activity of investors. We include daily return autocorrelation as the correlated trading patterns of investors contributes to daily return autocorrelations (Sias and Starks, 1997). The trading volume comovement measure or volume beta captures the sensitivity of individual stocks' trading volume to the aggregated market volume and captures correlated trading of investors (Chordia, Roll, and Subrahmanyam, 2000). In order to compute the volume beta for each stock in each quarter, we follow a similar procedure that Chordia, Roll, and Subrahmanyam (2000) and use a market model for volume and regress the percentage change in the daily turnover ratio for the individual stock on the concurrent percentage change in the market wide turnover (value-weighted average of all individual stock turnover, excluding the stock in the dependent variable).

We compute the quarterly values of all these proxies and report their statistics for both ASX200 stocks as well as non-ASX200 stocks in Table 9.

[Insert Table 9 about here]

Liquidity commonality for ASX200 stocks, on average, is higher than that of non-ASX200 stocks, as we expect. The average of commonality over the sample is 0.412 (0.085) for ASX200 (non-ASX200) stocks. The statistics of all proxies for correlated trading show that correlated trading in ASX200 stocks is more prevalent than in non-ASX200 stocks. For example, the average trading volume comovement measure for ASX200 stocks is more than 2.5 times higher than that for non-ASX200 stocks, and the autocorrelation for ASX200 stocks (−3.17%) is greater than that of non-ASX200 stocks (−8.388%).

[17] Results are available upon request.

We further split the non-ASX200 stocks into medium- and small-size stocks based on their market capitalizations at the end of the previous quarter.[18] Foreign institutional investors are likely to hold more medium stocks than small stocks in non-ASX200 constituents. The higher statistics of correlated trading measures for medium stocks compared to small stocks suggest that correlated trading happens more among medium-size stocks. For example, the mean of volume comovement measure (autocorrelation) is 0.377 (−6.413%) for the medium stocks, whereas it is 0.21 (−10.368%) for the small stocks. In summary, the results in Table 9 provide further support for our hypothesis that correlated trading is the channel through which foreign institutional investors impact commonality liquidity in Australia.

## 6. Further robustness tests

In this section, we run several further robustness tests for our findings that foreign institutional investors drive the liquidity commonality in Australia and correlated trading explains this relation.

In this study, we use value-weighted market liquidity to estimate the commonality in liquidity as in Eq. 2. It is important to show that our results are robust if we use equal-weighted portfolios. We repeat the calculation of liquidity beta for Table 5, which includes our main findings about the impact of foreign ownership on liquidity commonality. The results are reported in Appendix A (Table A1), and are qualitatively similar to those in Table 5.

Also, in the data preparation of our study and as a filtering procedure we exclude the stocks with a close price less than 1 cent. We also include the stocks that have a minimum of 25 valid observations per quarter. We recalculate liquidity beta and repeat the analysis for Table 5 with different price limits of 2 and 5 cents and a minimum 40 valid observations per quarter. The results presented in Table A2 in the Appendix are qualitatively similar to those in Table 5: foreign (local) ownership has a significant positive (insignificant) association with commonality.

Moreover, in Table 5, we use Amihud (2002) illiquidity measure as a control variable. In Table A3 in the Appendix, we show that our results in Table 5 are robust to the Pastor and Stambaugh (2003) illiquidity measure and quoted spread as alternative measures for the illiquidity level.

[18] The ASX200 stocks are considered as the large stocks in our sample.

Furthermore, Deng, Li and Li (2018) and Moshirian, Qian, Wee and Zhang (2017) use several firm characteristic variables in addition to size and illiquidity in their regression analysis to identify firm-level determinants of commonality in liquidity. As a robustness test, we test if our results in Table 5 are robust to additional firm characteristic control variables including book-to-market ratio, standard deviation of daily return over a quarter as a proxy for volatility, dividend yield and the log difference of stock prices. The results are reported in Table A4 in the Appendix and are qualitatively similar to our results in Table 5.

Finally, as we mentioned in Section 5, Sias and Starks (1997) suggest that daily return autocorrelation, as the correlated trading patterns of investors, contributes to daily return autocorrelations. Therefore, as a robustness test for our evidence of correlated trading among foreign investors, we take daily return autocorrelation, as a proxy for correlated trading, and examine the relation between return autocorrelation and foreign ownership in the cross-section. The results are reported in Table A5 in the Appendix. There is a significant positive relationship between return autocorrelation and foreign ownership in the univariate model as well as multivariate models when we include controls including local institutional ownership. These results indicate that correlated trading increases with the growth of foreign ownership and provides further support that foreign investors impact the commonality in liquidity through correlated trading.

## 7. Conclusion

We investigate the role of foreign and local institutional investors in liquidity commonality using a unique holding dataset from the Australian share market. We find that liquidity commonality increases with foreign institutional ownership across all stocks as well as large- and mid-cap stocks in the cross-section, whereas for local institutions, we find the positive relationship only in large-cap stocks. We further show that, over time, commonality in liquidity is higher for large firms compared to small firms. The spread of the commonality in liquidity between large and small firms increases after the year 2001 and is explained by the difference in their fraction of foreign institutional ownership. Our findings provide further support for previous studies that show institutional investing is a determinant of commonality in stock liquidity (Koch, Ruenzi, and Starks, 2016). However, we provide evidence that foreign, rather than local, investors are the main driver of commonality in liquidity. We provide evidence for correlated trading by both local and foreign institutions that can explain the positive relationship between fraction of ownership and commonality in liquidity.

Our findings suggest that a distinction should be made between local institutions and foreign institutions in their impact on liquidity commonality. Our results suggest that in Australia the exposure of large stocks to systematic liquidity risk has increased since the year 2001, which implies that the ability to diversify liquidity shocks by holding large (and relatively liquid) stocks has declined.

**Appendix A**

See Tables A1-A5

**Table A1**

Market liquidity beta estimated with an equal-weighted market portfolio and stock ownership in the cross-section. We estimate $\beta_L$ in a regression similar to Eq. 2, in which we use an equal-weighted market portfolio to compute changes in the daily illiquidity measure, and repeat Table 5. Numbers in brackets are *t*-statistics, adjusted for heteroskedasticity and autocorrelation using Newey and West (1987) standard errors with two lags.

| Variables | [1] | [2] | [3] | [4] | [5] |
|---|---|---|---|---|---|
| Intercept | 0.238 | -0.25 | -0.136 | -0.109 | -0.137 |
| | [5.22] | [-3.17] | [-2.01] | [-1.24] | [-1.54] |
| Institutional Ownership | 0.577 | 0.309 | | | |
| | [8.79] | [5.28] | | | |
| Foreign Institutional Ownership | | | 0.427 | | 0.427 |
| | | | [7.08] | | [7.06] |
| Local Institutional Ownership | | | | -0.059 | 0.003 |
| | | | | [-0.57] | [0.03] |
| Size | | 0.108 | 0.096 | 0.116 | 0.096 |
| | | [10.16] | [8.80] | [10.10] | [8.42] |
| Illiq | | 0.004 | 0.004 | 0.002 | 0.004 |
| | | [0.86] | [0.78] | [0.33] | [0.82] |

**Table A2**
Market liquidity beta and stock ownership in the cross-section with alternative sample filters. We repeat Table 5 with the following filters applied to the sample. Panel A excludes stocks with a price lower than 2 cents from the sample. Panel B excludes stocks with a price lower than 5 cents from the sample. Panel C retains firm-quarters in the sample with a minimum of 40 observations. Numbers in brackets are *t*-statistics, adjusted for heteroskedasticity and autocorrelation using Newey and West (1987) standard errors with two lags.

| Panel A: 2 cents minimum price | | | | | |
|---|---|---|---|---|---|
| Variables | [1] | [2] | [3] | [4] | [5] |
| Intercept | -0.004 | -0.44 | -0.398 | -0.411 | -0.421 |
| | [-0.26] | [-20.64] | [-19.93] | [-13.74] | [-14.04] |
| Institutional Ownership | 0.317 | 0.12 | | | |
| | [8.69] | [3.70] | | | |
| Foreign Institutional Ownership | | | 0.136 | | 0.142 |
| | | | [4.08] | | [4.22] |
| Local Institutional Ownership | | | | 0.049 | 0.065 |
| | | | | [0.88] | [1.12] |
| Size | | 0.089 | 0.086 | 0.094 | 0.087 |
| | | [23.90] | [20.74] | [22.57] | [19.73] |
| Illiq | | 0.003 | 0.003 | 0.002 | 0.003 |
| | | [0.73] | [0.72] | [0.55] | [0.76] |

| Panel B: 5 cents minimum price | | | | | |
|---|---|---|---|---|---|
| Variables | [1] | [2] | [3] | [4] | [5] |
| Intercept | -0.008 | -0.521 | -0.46 | -0.481 | -0.496 |
| | [-0.45] | [-20.57] | [-19.88] | [-14.27] | [-15.11] |
| Institutional Ownership | 0.347 | 0.166 | | | |
| | [10.41] | [4.97] | | | |
| Foreign Institutional Ownership | | | 0.186 | | 0.193 |
| | | | [5.10] | | [5.21] |
| Local Institutional Ownership | | | | 0.073 | 0.096 |
| | | | | [1.48] | [1.89] |
| Size | | 0.096 | 0.092 | 0.102 | 0.093 |
| | | [25.28] | [21.02] | [24.10] | [20.21] |
| Illiq | | 0.011 | 0.01 | 0.009 | 0.011 |
| | | [2.16] | [2.11] | [1.98] | [2.14] |

*To be continued in next page*

**Table A2** (*Continued*)

| Panel C: Firm-quarters with minimum 40 observations | | | | | |
|---|---|---|---|---|---|
| Variables | [1] | [2] | [3] | [4] | [5] |
| Intercept | 0.03 | -0.486 | -0.443 | -0.451 | -0.471 |
| | [2.04] | [-15.71] | [-17.25] | [-12.62] | [-12.63] |
| Institutional Ownership | 0.296 | 0.119 | | | |
| | [10.40] | [4.81] | | | |
| Foreign Institutional Ownership | | | 0.124 | | 0.129 |
| | | | [4.71] | | [4.95] |
| Local Institutional Ownership | | | | 0.058 | 0.081 |
| | | | | [1.02] | [1.42] |
| Size | | 0.096 | 0.093 | 0.1 | 0.094 |
| | | [29.13] | [26.39] | [24.63] | [22.89] |
| Illiq | | 0.008 | 0.007 | 0.004 | 0.007 |
| | | [1.25] | [1.18] | [0.65] | [1.10] |

**Table A3**

Market liquidity beta and stock ownership in the cross-section with alternative illiquidity controls. We repeat Table 5 with alternative illiquidity measures as a control. Panel A replaces the daily Amihud (2002) illiquidity measure with the Pastor and Stambaugh (2003) illiquidity measure as a control. Panel B replaces the daily Amihud (2002) illiquidity measure with the quoted percentage spread as a control. Numbers in brackets are *t*-statistics, adjusted for heteroskedasticity and autocorrelation using Newey and West (1987) standard errors with two lags.

| Panel A: Pastor-Stambaugh (2003) illiquidity as a control | | | | |
|---|---|---|---|---|
| Variables | [1] | [2] | [3] | [4] |
| Intercept | -0.465 | -0.421 | -0.435 | -0.445 |
| | [-13.78] | [-14.57] | [-11.60] | [-11.95] |
| Institutional Ownership | 0.132 | | | |
| | [3.67] | | | |
| Foreign Institutional Ownership | | 0.144 | | 0.149 |
| | | [3.74] | | [3.88] |
| Local Institutional Ownership | | | 0.058 | 0.081 |
| | | | [0.83] | [1.16] |
| Size | 0.093 | 0.09 | 0.098 | 0.091 |
| | [19.70] | [18.83] | [18.71] | [18.24] |
| Pastor-Stambaugh Illiquidity | 0.146 | 0.142 | 0.132 | 0.15 |
| | [1.07] | [1.05] | [0.95] | [1.10] |

| Panel B: Percentage spread as a control | | | | |
|---|---|---|---|---|
| Variables | [1] | [2] | [3] | [4] |
| Intercept | -0.465 | -0.445 | -0.434 | -1.569 |
| | [-13.78] | [-11.95] | [-12.71] | [-5.50] |
| Institutional Ownership | 0.098 | | | |
| | [2.93] | | | |
| Foreign Institutional Ownership | | 0.105 | | 0.109 |
| | | [2.90] | | [3.03] |
| Local Institutional Ownership | | | 0.047 | 0.065 |
| | | | [0.72] | [0.98] |
| Size | 0.095 | 0.093 | 0.099 | 0.094 |
| | [20.06] | [19.63] | [19.07] | [19.05] |
| Percentage Spread | -1.574 | -1.577 | -1.747 | -1.569 |
| | [-5.64] | [-5.61] | [-6.02] | [-5.50] |

**Table A4**

Market liquidity beta and stock ownership in the cross-section with additional controls. We repeat Table 5 and control for additional firm-specific variables. *BM* is the natural log of book to market ratio, computed as one plus the ratio of the book equity at the end of last June to the market capitalization at the end of last December. *DY* is the dividend yield, computed as the ratio of total cash dividend payments during the last calendar year to the stock price at the end of last December. *STDRET* is the standard deviation of daily returns over the previous quarter. *RE* is the log difference of stock prices between the beginning and the end of the previous quarter. Numbers in brackets are *t*-statistics, adjusted for heteroskedasticity and autocorrelation using Newey and West (1987) standard errors with two lags.

| Variables | [1] | [2] | [3] | [4] |
|---|---|---|---|---|
| Intercept | -0.666 | -0.617 | -0.648 | -0.656 |
| | [-11.79] | [-12.67] | [-11.27] | [-11.33] |
| Institutional Ownership | 0.125 | | 0.098 | |
| | [3.42] | | [2.93] | |
| Foreign Institutional Ownership | | 0.13 | | 0.135 |
| | | [3.26] | | [3.43] |
| Local Institutional Ownership | | | 0.078 | 0.095 |
| | | | [1.25] | [1.52] |
| Size | 0.111 | 0.108 | 0.116 | 0.11 |
| | [19.18] | [18.89] | [18.47] | [18.52] |
| Illiq | 0.002 | 0.002 | 0 | 0.002 |
| | [0.46] | [0.49] | [0.08] | [0.51] |
| BM | 0.002 | 0.002 | 0.001 | 0.002 |
| | [0.44] | [0.33] | [0.22] | [0.41] |
| DY | 0.162 | 0.186 | 0.156 | 0.174 |
| | [0.62] | [0.68] | [0.59] | [0.66] |
| STDRET | 2.508 | 2.402 | 2.559 | 2.493 |
| | [4.54] | [4.42] | [4.58] | [4.38] |
| RE | -0.099 | -0.095 | -0.101 | -0.097 |
| | [-3.36] | [-3.15] | [-3.35] | [-3.16] |

**Table A5**

Correlated trading: return autocorrelation and foreign institutional ownership. This table reports estimates from Fama and Macbeth (1973) specification of the following regression:

$$Autocorr_{i,t} = \alpha + b_1 Fown_{i,t-1} + \delta Controls_{i,t-1} + v_{i,t}$$

where $Autocorr_{i,t}$ is the daily return autocorrelation, estimated for firm *i* over quarter *t*. $Fown_{i,t-1}$ is the foreign institutional ownership and computed as the number of shares held by foreign institutional investors divided by the total shares outstanding. Control variables include Size, the natural logarithm of the firm's market capitalization (in millions); Illiq, the daily average of the firm's Amihud (2002) illiquidity measure and Local Institutional Ownership, the number of shares held by local institutional investors divided by the total shares outstanding. All control variables are computed at the end of the previous quarter. We present the time-series averages of the coefficients. Numbers in brackets are *t*-statistics, adjusted for heteroskedasticity and autocorrelation using Newey and West (1987) standard errors with two lags.

| Variables | [1] | [2] | [3] | [4] |
|---|---|---|---|---|
| Intercept | -0.105 | -0.14 | -0.132 | -0.119 |
| | [-14.40] | [-12.51] | [-12.04] | [-8.26] |
| Foreign Institutional Ownership | 0.133 | 0.1 | 0.098 | 0.095 |
| | [10.78] | [9.10] | [9.13] | [8.68] |
| Size | | 0.008 | 0.007 | 0.006 |
| | | [5.55] | [4.79] | [4.10] |
| Illiq | | | -0.003 | -0.003 |
| | | | [-3.27] | [-3.38] |
| Local Institutional Ownership | | | | -0.034 |
| | | | | [-2.13] |

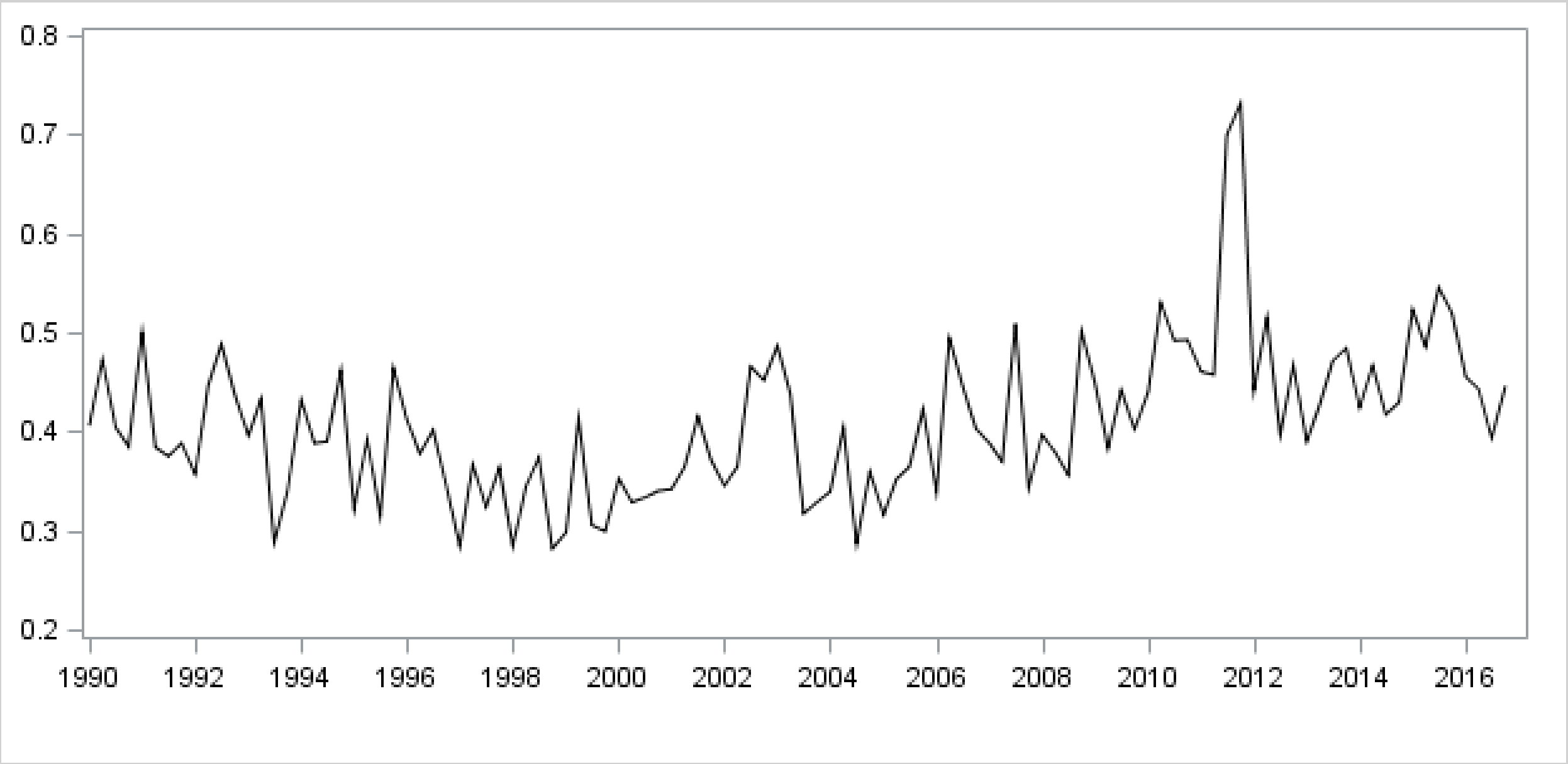

**Fig. 1.** Quarterly standard deviation of changes in market illiquidity. We compute $\Delta illiq_{i,d}$ as the logarithmic change in daily Amihud (2002) illiquidity measure for each firm *i*. $\Delta illiq_{mkt,d}$ is the daily value-weighted average of $\Delta illiq_{i,d}$ for all stocks in our sample from January 1990 to December 2016. We present the quarterly standard deviation of $\Delta illiq_{mkt,d}$.

a)

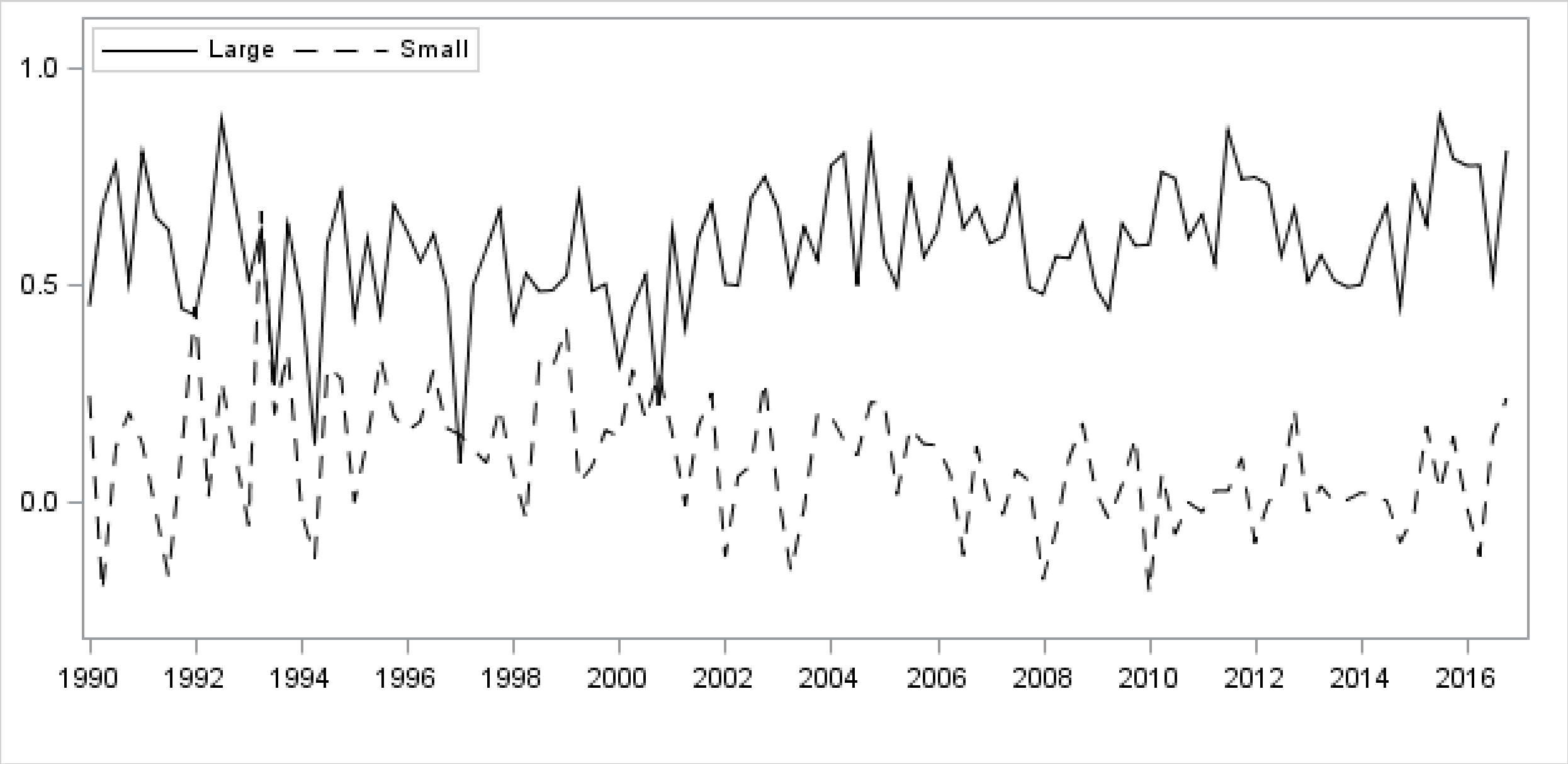

Market liquidity beta for large and small firms

b)

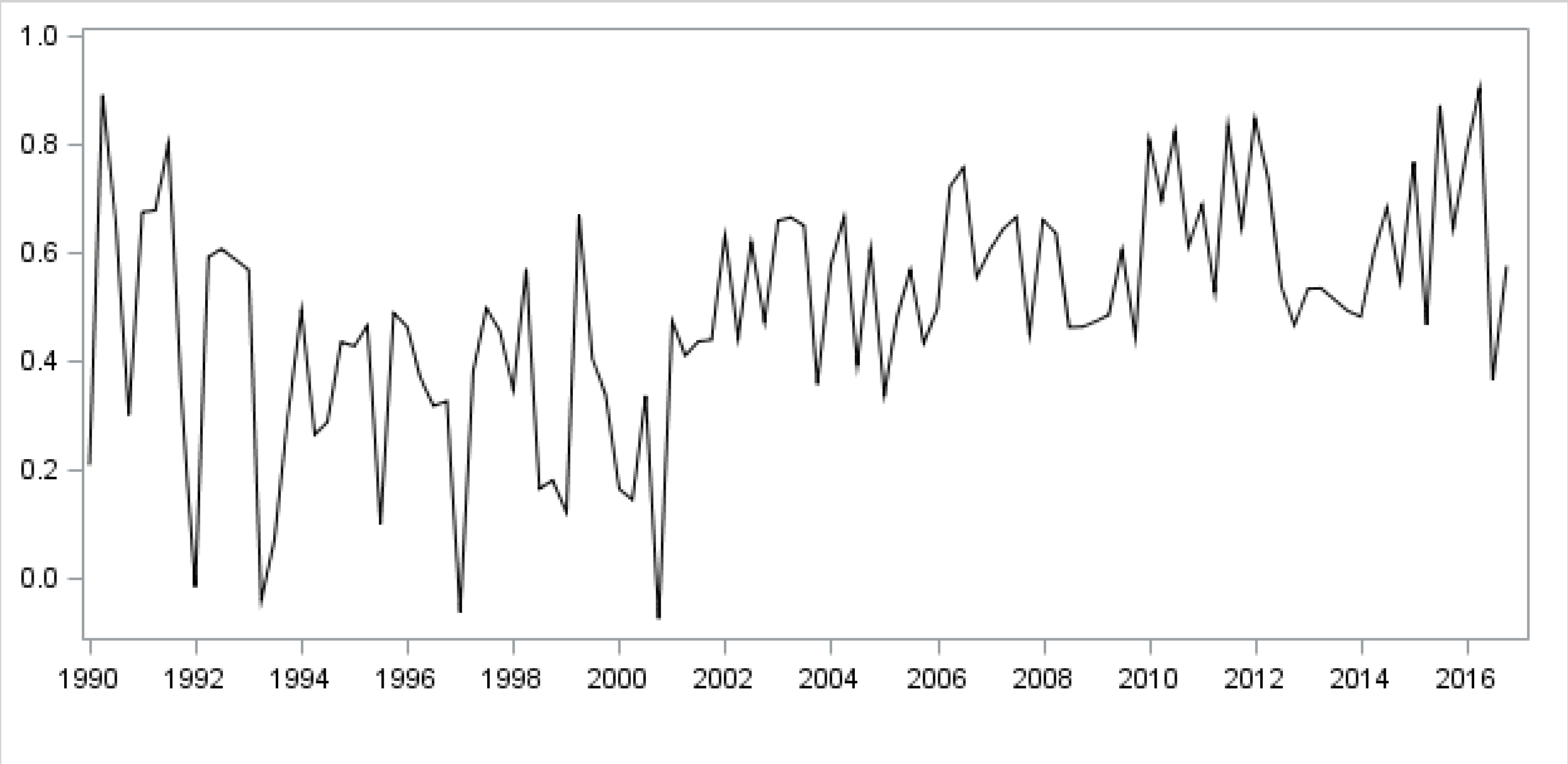

Difference in market liquidity beta between large and small firms

**Fig. 2.** Time series of market liquidity beta. In each quarter and for each firm, we regress the firm's daily changes in the Amihud (2002) illiquidity measure on the daily changes in the value-weighted illiquidity measure of the market portfolio and a set of controls. For each regression, we remove firm *i* from the market portfolio in computing the daily changes in value-weighted illiquidity. For each quarter, we retain firms with at least 25 valid observations. We sort firms into terciles each quarter based on the market capitalization at the end of the previous quarter: small (bottom tercile), mid-cap (middle tercile) or large (top tercile). Panel (a) plots the quarterly cross-sectional average of $\beta_L$ for large and small firms, and Panel (b) presents the difference in average $\beta_L$ between large and small firms.

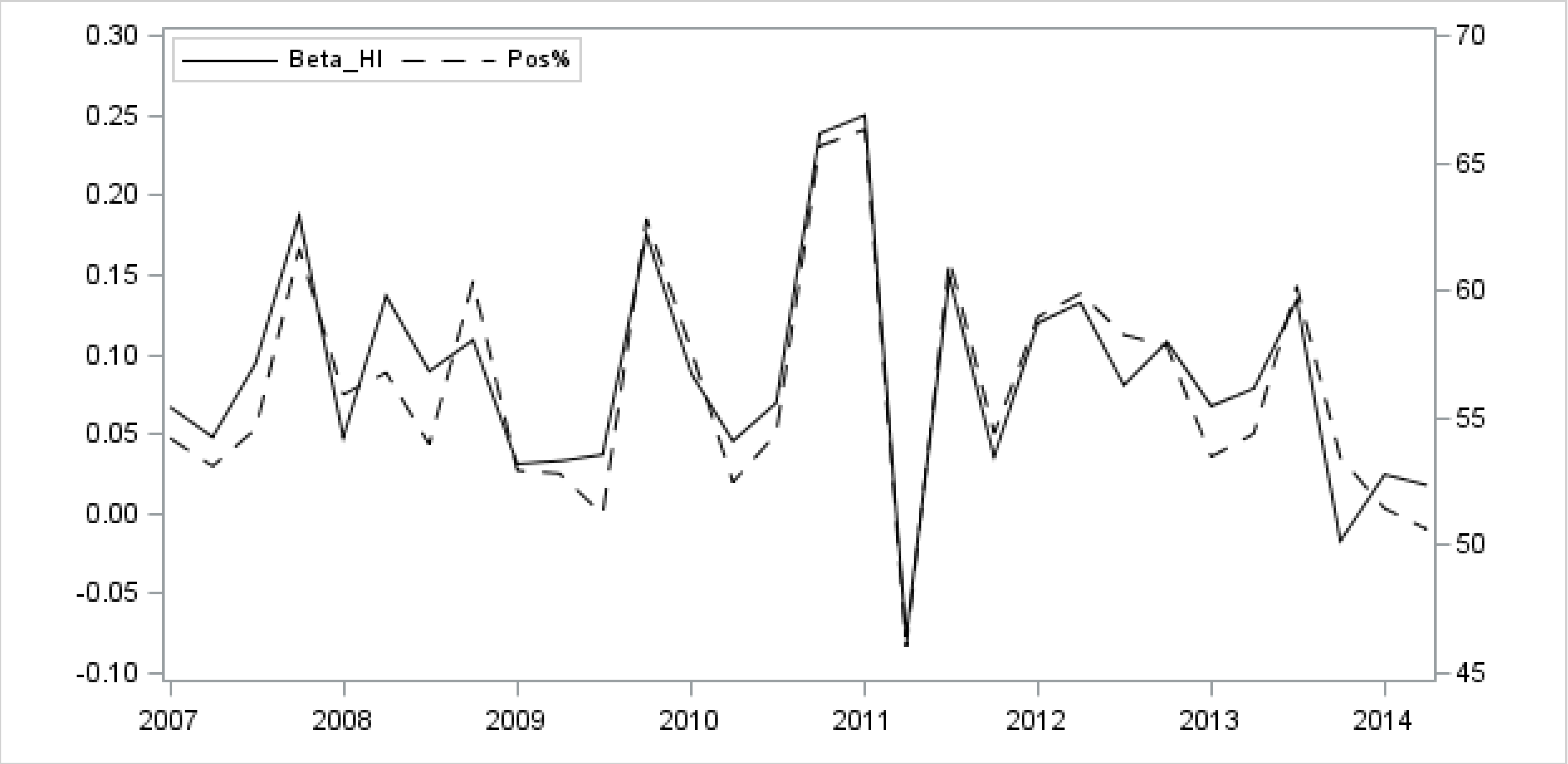

**Fig. 3.** Time series of high ownership liquidity beta for foreign institutional investors. The high foreign institutional ownership liquidity beta ($\beta_{HI,i,t}$ ) for stock *i* is estimated as in Eq. 7. We plot the time series of cross-sectional means of $\beta_{HI}$ and the percentage of positive $\beta_{HI}$ among all sample stocks.

**Table 1**

Summary Statistics. This table reports summary statistics for the sample stocks trading on the ASX at any point between 1 January1990 and 31 December 2016. Size is the market capitalization of the stock's equity at the end of the previous quarter. Price is the average close price of the stock over the quarter. Dollar Volume is the average daily dollar volume over the quarter. We sort firms into terciles each quarter based on the market capitalization at the end of the previous quarter: small (bottom tercile), mid-cap (middle tercile) or large (top tercile). Panel A reports means, medians, standard deviations, minimums and maximums for the full sample of stock quarters. Panel B reports the summary statistics on ownership percentage by each investor category. For each quarter, we compute ownership percentages as the number of shares held by each investor category divided by the total number of shares outstanding at the end of the previous quarter between 2007 and 2014. Panel C reports the summary statistics on changes in stock ownership. We compute the dollar value of shareholdings by multiplying the number of shares held by each investor type with the close price. We then aggregate the holdings across the sample stocks to compute total dollar shareholdings by each investor category. We present the daily average dollar shareholdings and ownership percentage for the year 2007 and the year 2014, as well as changes in dollar and percentage shareholdings. We delete daily observations from the sample if the stock price crossed a minimum tick level during the day in either direction. Stock price must be equal or greater than $0.01 to be included in the sample. We retain stock quarters with at least 25 valid observations from the sample remaining after applying previous filters.

| Panel A: Summary statistics | | | | | | |
|---|---|---|---|---|---|---|
| | | Mean | Median | Std Dev | Min | Max |
| All Stocks | Size ($millions) | 1816.770 | 220.380 | 7327.290 | 0.099 | 155931.960 |
| | Price ($) | 3.879 | 1.469 | 8.095 | 0.012 | 180.982 |
| | Dollar Volume ($thousands) | 5603.510 | 317.022 | 24804.500 | 2.888 | 781128.970 |
| Large | Size ($millions) | 5153.030 | 1676.780 | 12017.440 | 283.384 | 155931.960 |
| | Price ($) | 8.830 | 4.871 | 12.302 | 0.066 | 180.982 |
| | Dollar Volume ($thousands) | 16125.070 | 4033.140 | 40987.290 | 3.691 | 781128.970 |
| Mid-Cap | Size ($millions) | 252.177 | 218.528 | 145.770 | 37.307 | 1279.580 |
| | Price ($) | 2.178 | 1.487 | 2.546 | 0.038 | 48.896 |
| | Dollar Volume ($thousands) | 551.954 | 299.468 | 780.330 | 6.803 | 15406.390 |
| Small | Size ($millions) | 47.122 | 37.544 | 37.803 | 0.099 | 397.846 |
| | Price ($) | 0.631 | 0.374 | 0.944 | 0.012 | 39.088 |
| | Dollar Volume ($thousands) | 140.418 | 74.202 | 345.551 | 2.888 | 21241.920 |
| Panel B: Ownership percentage | | | | | | |
| | | Mean | Median | Std Dev | Min | Max |
| All Stocks | Institutions (in %) | 56.05 | 57.21 | 20.60 | 0.14 | 99.67 |
| | Foreign Institutions (in %) | 31.43 | 30.04 | 19.62 | 0.03 | 92.29 |
| | Local Institutions (in %) | 24.62 | 23.92 | 11.84 | 0.01 | 78.00 |
| Large | Institutions (in %) | 62.00 | 66.72 | 22.06 | 0.14 | 99.67 |
| | Foreign Institutions (in %) | 41.83 | 43.52 | 18.88 | 0.04 | 87.43 |
| | Local Institutions (in %) | 20.17 | 20.10 | 9.94 | 0.01 | 76.18 |

*To be continued in next page*

**Table 1** *(Continued)*

| | | Mean | Median | Std Dev | Min | Max |
|---|---|---|---|---|---|---|
| Mid-Cap | Institutions (in %) | 55.88 | 58.17 | 19.84 | 1.17 | 98.26 |
| | Foreign Institutions (in %) | 30.48 | 29.66 | 18.13 | 0.03 | 92.29 |
| | Local Institutions (in %) | 25.41 | 24.50 | 12.33 | 0.17 | 78.00 |
| Small | Institutions (in %) | 48.50 | 48.21 | 16.68 | 1.41 | 97.26 |
| | Foreign Institutions (in %) | 19.06 | 15.84 | 14.04 | 0.09 | 88.61 |
| | Local Institutions (in %) | 29.45 | 29.08 | 11.39 | 0.75 | 75.79 |

Panel C: Changes in shareholdings

| | Avg in 2007 (in %) | Avg in 2014 (in %) | Change in Percent Holdings (in %) | Avg in 2007 (in Billion $) | Avg in 2014 (in Billion $) | Change in Dollar Holdings (in %) |
|---|---|---|---|---|---|---|
| Institutions | 49.83 | 61.61 | 11.77 | 760.738 | 971.601 | 27.72 |
| Foreign Institutions | 26.08 | 33.99 | 7.91 | 511.318 | 627.935 | 22.81 |
| Local Institutions | 23.76 | 27.62 | 3.86 | 249.419 | 343.666 | 37.79 |

**Table 2**
Means of market liquidity beta. In each quarter and for each firm, we regress the firm's daily changes in the Amihud (2002) illiquidity measure on the daily changes in the value-weighted illiquidity measure of the market portfolio:

$$\Delta illiq_{i,d} = \alpha_i + \beta_{L,i} \, \Delta illiq_{mkt,d} + \sum_{j=1}^{J} \delta_j \, Z_{ji,d} + \epsilon_{i,d}$$

For each regression, we remove firm *i* from the market portfolio in computing $\Delta illiq_{mkt,d}$. We control for the variables ($Z_{ji,d}$) which may correlate with the firm's liquidity: lead and lag changes in the market portfolio; lead, lag, and contemporaneous market returns; and contemporaneous squared stock return. For each quarter, we retain firms with at least 25 valid observations. We sort firms into terciles each quarter based on the market capitalization at the end of the previous quarter: small (bottom tercile), mid-cap (middle tercile) or large (top tercile). For each firm year, we compute the mean of quarterly betas. We report annual cross-sectional average of these means as well as *t*-statistics of $\beta_L$ and the percentage that are positive. We also report the difference in means and *t*-statistics of $\beta_L$ between large and small firms.

| | All | | | Large | | | Small | | | Large minus Small | |
|---|---|---|---|---|---|---|---|---|---|---|---|
| Year | Mean | t-stat | % Pos | Mean | t-stat | % Pos | Mean | t-stat | % Pos | Mean | t-stat |
| 1990 | 0.154 | [2.08] | 60 | 0.191 | [1.86] | 73 | 0.019 | [0.13] | 44 | 0.172 | [0.87] |
| 1991 | 0.272 | [3.72] | 64 | 0.524 | [5.29] | 79 | 0.117 | [1] | 58 | 0.406 | [2.47] |
| 1992 | 0.186 | [2.4] | 60 | 0.483 | [4.69] | 76 | -0.035 | [-0.27] | 51 | 0.518 | [2.85] |
| 1993 | 0.298 | [6.08] | 68 | 0.331 | [4.66] | 77 | 0.341 | [4.24] | 68 | -0.01 | [-0.09] |
| 1994 | 0.188 | [3.65] | 64 | 0.296 | [3.17] | 76 | 0.077 | [0.99] | 55 | 0.219 | [1.79] |
| 1995 | 0.231 | [3.82] | 60 | 0.328 | [3.61] | 68 | 0.171 | [1.85] | 57 | 0.157 | [1.17] |
| 1996 | 0.182 | [4.44] | 61 | 0.314 | [4.82] | 70 | 0.112 | [1.58] | 59 | 0.201 | [2] |
| 1997 | 0.247 | [4.24] | 57 | 0.374 | [3.78] | 60 | 0.129 | [1.44] | 54 | 0.245 | [1.82] |
| 1998 | 0.279 | [5.44] | 61 | 0.434 | [5.91] | 69 | 0.1 | [1.13] | 52 | 0.334 | [2.76] |
| 1999 | 0.261 | [6.88] | 62 | 0.295 | [4.74] | 68 | 0.199 | [3.1] | 57 | 0.095 | [1.04] |
| 2000 | 0.161 | [3.68] | 57 | 0.269 | [3.67] | 63 | 0.07 | [0.95] | 53 | 0.2 | [1.91] |
| 2001 | 0.267 | [6.3] | 63 | 0.429 | [7.27] | 75 | 0.076 | [0.97] | 51 | 0.353 | [3.42] |
| 2002 | 0.092 | [2.23] | 54 | 0.234 | [3.48] | 61 | 0.064 | [0.87] | 52 | 0.17 | [1.68] |
| 2003 | 0.174 | [4.75] | 62 | 0.349 | [7.32] | 73 | 0.044 | [0.62] | 54 | 0.305 | [3.48] |
| 2004 | 0.289 | [8.46] | 64 | 0.486 | [9.03] | 76 | 0.177 | [2.83] | 54 | 0.309 | [3.64] |
| 2005 | 0.181 | [4.84] | 57 | 0.392 | [7.57] | 71 | 0.115 | [1.71] | 51 | 0.277 | [3.08] |

*To be continued in next page*

**Table 2** *(Continued)*

| | | | | | | | | | | | | |
|---|---|---|---|---|---|---|---|---|---|---|---|---|
| 2006 | 0.14 | [5.44] | 60 | 0.333 | [8.57] | 74 | 0.07 | [1.45] | 56 | 0.263 | [4.13] |
| 2007 | 0.074 | [2.86] | 56 | 0.278 | [7.24] | 70 | -0.064 | [-1.34] | 47 | 0.342 | [5.52] |
| 2008 | 0.032 | [1.02] | 56 | 0.28 | [6.52] | 72 | -0.16 | [-2.71] | 46 | 0.441 | [5.99] |
| 2009 | 0.105 | [3.79] | 58 | 0.23 | [5.69] | 67 | 0.058 | [1.04] | 52 | 0.172 | [2.47] |
| 2010 | 0.043 | [1.5] | 57 | 0.251 | [6.34] | 71 | -0.121 | [-2.25] | 48 | 0.372 | [5.34] |
| 2011 | 0.142 | [6.12] | 62 | 0.334 | [9.55] | 79 | 0.006 | [0.16] | 50 | 0.327 | [6.02] |
| 2012 | 0.084 | [3.56] | 56 | 0.322 | [9.83] | 79 | -0.071 | [-1.75] | 43 | 0.392 | [7.23] |
| 2013 | 0.068 | [2.25] | 56 | 0.136 | [3] | 63 | 0.081 | [1.46] | 56 | 0.055 | [0.75] |
| 2014 | 0.077 | [2.69] | 56 | 0.288 | [7.16] | 70 | -0.041 | [-0.77] | 48 | 0.329 | [4.73] |
| 2015 | 0.092 | [3.99] | 59 | 0.287 | [8.32] | 74 | -0.007 | [-0.16] | 52 | 0.293 | [5.23] |
| 2016 | 0.145 | [6.38] | 62 | 0.318 | [10.22] | 76 | 0.09 | [2.1] | 56 | 0.228 | [4.08] |

**Table 3**
Unit-root tests. Panel A reports the estimates of Dickey-Fuller (1981) unit-root tests for the time series of quarterly average market liquidity betas of large and small firms as well as the difference between large and small firms. For each time series, we run the following regression:

$$\beta_{L,t} = \alpha + \rho\beta_{L,t-1} + \gamma t + \upsilon_t$$

where $\alpha$ is a drift, $t$ is a time trend and $\beta_{L,t-1}$ is the first lag of the average market liquidity beta at time $t$. Panel A presents the estimates on $\beta_{L,t-1}$ and p-value for the full sample period 1990–2016 as well as for two sub-sample periods: 1990–2000 and 2001–2016. Panel B reports estimates from the following time-series regression:

$$\beta_{L,t} = \alpha + \gamma t + \upsilon_t$$

where $\alpha$ is a constant and $t$ is a time trend. Panel B presents the estimates on $\alpha$ and $t$ and respective p-value for the full sample period 1990–2016 as well as for the two sub-sample periods: 1990–2000 and 2001–2016. The p-values are computed adjusting for heteroskedasticity and autocorrelation using Newey and West (1987) standard errors with two lags.

| Panel A: Stochastic time trend | | | | | | | |
|---|---|---|---|---|---|---|---|
| Variable | Year | 1990-2016 | | 1990-2000 | | 2001-2016 | |
| | Size Group | Estimate | p-value | Estimate | p-value | Estimate | p-value |
| $\beta_{L,t-1}$ | Large | -0.81 | <0.001 | -1.209 | <0.001 | -0.799 | <0.001 |
| | Small | -1.126 | <0.001 | -1.327 | <0.001 | -1.172 | <0.001 |
| | Large − Small | -0.726 | <0.001 | -1.128 | <0.001 | -0.769 | <0.001 |

| Panel B: Deterministic time trend | | | | | | | |
|---|---|---|---|---|---|---|---|
| Variable | | 1990-2016 | | 1990-2000 | | 2001-2016 | |
| | Size Group | Estimate | p-value | Estimate | p-value | Estimate | p-value |
| Constant | Large | 0.52 | <0.001 | 0.626 | <0.001 | 0.549 | <0.001 |
| | Small | 0.184 | <0.001 | 0.106 | 0.056 | 0.145 | 0.035 |
| | Large − Small | 0.337 | <0.001 | 0.52 | <0.001 | 0.404 | <0.001 |
| | Size Group | Estimate | p-value | Estimate | p-value | Estimate | p-value |
| Time Trend | Large | 0.001 | 0.012 | -0.004 | 0.012 | 0.001 | 0.215 |
| | Small | -0.002 | 0.001 | 0.003 | 0.167 | -0.001 | 0.131 |
| | Large − Small | 0.003 | <0.001 | -0.007 | 0.018 | 0.002 | 0.01 |

**Table 4**
Implied trading activity by institutional investor groups. Each quarter, we sort stocks into quintiles based on the fraction of ownership by foreign institutions and local institutions at the end of the previous quarter, and compute average daily turnover as volume divided by shares outstanding for stocks in each ownership quintile. Panel A reports time-series means of average daily turnover (in percentage) for each ownership quintile. Panel B reports ownership-weighted turnover and holdings periods computed as follows. First, we compute average daily turnover (in percentage) as volume divided by shares outstanding for stocks over a quarter. Then, for each quarter, we compute the cross-sectional average of turnover, weighted by ownership percentage, at the end of the previous quarter for both investor groups, and compute the cross-sectional holding period (in days) as the inverse of cross-sectional average of turnover. We report time series means of cross-sectional averages of turnover and holding period for each investor group. Panel C presents average of daily turnover and holding period of ASX200 stocks and non-ASX200 stocks. *t*-statistics are in brackets.

| Panel A: Average turnover (%) by ownership | | |
|---|---|---|
| Ownership Quintile | Foreign Institutions | Local Institutions |
| Lo | 0.184 | 0.232 |
| 2 | 0.247 | 0.331 |
| 3 | 0.279 | 0.358 |
| 4 | 0.351 | 0.327 |
| Hi | 0.463 | 0.277 |
| Hi - Lo | 0.279 | 0.045 |
| *t*-stat | [22.3] | [5.6] |

| Panel B: Turnover and holding period by investor group | | | |
|---|---|---|---|
| | Foreign Institutions | Local Institutions | Foreign - Local |
| Turnover (%) | 0.365 | 0.309 | 0.057 |
| | | | [16.48] |
| Holding period | 276 | 328 | -53 |
| | | | [-12.88] |

| Panel C: Turnover and holding period of ASX200 and non-ASX200 stocks | | | |
|---|---|---|---|
| | ASX200 | Non-ASX200 | Diff |
| Turnover (%) | 0.457 | 0.23 | 0.227 |
| | | | [21.12] |
| Holding period | 221 | 451 | -229 |
| | | | [-15.31] |

**Table 5**

Market liquidity beta and stock ownership in the cross-section. This table reports estimates from Fama and Macbeth (1973) specification of the following regression:

$$\beta_{L,i,t} = \alpha + b_1 Own_{i,t-1} + b_2 Size_{i,t-1} + b_3 Illiq_{i,t-1} + \upsilon_{i,t}$$

where $\beta_{L,i,t}$ is estimated as in Eq. 2. $Own_{i,t-1}$ is the institutional ownership computed as the number of shares held by the investor type divided by the total shares outstanding for the following investor categories: institutions, foreign institutions and local institutions. *Size* is the natural logarithm of the firm's market capitalization (in millions). $Illiq_{i,t-1}$ is the daily average of the firm's Amihud (2002) illiquidity measure, computed over the previous quarter. $Size_{i,t-1}$ and $Own_{i,t-1}$ are computed at the end of the previous quarter. We sort firms into terciles each quarter based on the market capitalization at the end of the previous quarter: small (bottom tercile), mid-cap (middle tercile) or large (top tercile). We present the time-series averages of the coefficients for all stocks as well as for each size group. Numbers in brackets are *t*-statistics, adjusted for heteroskedasticity and autocorrelation using Newey and West (1987) standard errors with two lags.

| Size Group | Variables | [1] | [2] | [3] | [4] | [5] |
|---|---|---|---|---|---|---|
| All Stocks | Intercept | -0.015 | -0.403 | -0.367 | -0.38 | -0.388 |
| | | [-1.02] | [-15.39] | [-15.69] | [-12.94] | [-13.04] |
| | Institutional Ownership | 0.315 | 0.099 | | | |
| | | [9.44] | [3.43] | | | |
| | Foreign Institutional Ownership | | | 0.11 | | 0.116 |
| | | | | [3.89] | | [3.99] |
| | Local Institutional Ownership | | | | 0.04 | 0.056 |
| | | | | | [0.80] | [1.08] |
| | Size | | 0.086 | 0.084 | 0.09 | 0.085 |
| | | | [21.11] | [20.86] | [20.02] | [20.19] |
| | Illiq | | 0.005 | 0.005 | 0.004 | 0.005 |
| | | | [1.15] | [1.14] | [0.99] | [1.15] |
| Large | Intercept | 0.179 | -0.574 | -0.542 | -0.546 | -0.577 |
| | | [6.56] | [-11.09 | [-9.96] | [-10.63] | [-11.58] |
| | Institutional Ownership | 0.256 | 0.137 | | | |
| | | [7.39] | [3.51] | | | |
| | Foreign Institutional Ownership | | | 0.141 | | 0.139 |
| | | | | [2.93] | | [2.99] |
| | Local Institutional Ownership | | | | 0.163 | 0.145 |
| | | | | | [2.32] | [2.24] |
| | Size | | 0.107 | 0.107 | 0.111 | 0.108 |
| | | | [17.29] | [16.79] | [20.17] | [17.51] |
| | Illiq | | -0.417 | -0.45 | -0.471 | -0.417 |
| | | | [-2.33] | [-2.36] | [-2.36] | [-2.34] |

*To be continued in next page*

**Table 5** *(Continued)*

| | | | | | | |
|---|---|---|---|---|---|---|
| Mid-Cap | Intercept | 0.039 | -0.339 | -0.314 | -0.321 | -0.293 |
| | | [1.33] | [-2.45] | [-2.40] | [-2.28] | [-2.04] |
| | Institutional Ownership | 0.094 | 0.056 | | | |
| | | [1.69] | [0.92] | | | |
| | Foreign Institutional Ownership | | | 0.102 | | 0.092 |
| | | | | [2.06] | | [1.55] |
| | Local Institutional Ownership | | | | -0.046 | -0.035 |
| | | | | | [-0.72] | [-0.45] |
| | Size | | 0.074 | 0.069 | 0.078 | 0.068 |
| | | | [3.15] | [2.92] | [3.26] | [2.85] |
| | Illiq | | -0.027 | -0.021 | -0.022 | -0.028 |
| | | | [-0.80] | [-0.70] | [-0.74] | [-0.81] |
| Small | Intercept | 0.001 | -0.056 | -0.048 | 0.015 | -0.009 |
| | | [0.03] | [-0.77] | [-0.62] | [0.21] | [-0.12] |
| | Institutional Ownership | 0.034 | 0.033 | | | |
| | | [0.50] | [0.47] | | | |
| | Foreign Institutional Ownership | | | 0.115 | | 0.113 |
| | | | | [1.41] | | [1.31] |
| | Local Institutional Ownership | | | | -0.126 | -0.092 |
| | | | | | [-1.50] | [-0.99] |
| | Size | | 0.015 | 0.011 | 0.01 | 0.009 |
| | | | [0.77] | [0.58] | [0.58] | [0.44] |
| | Illiq | | 0.001 | 0.001 | 0 | 0.001 |
| | | | [0.22] | [0.22] | [0.06] | [0.10] |

**Table 6**

Spread in market liquidity beta and difference in stock ownership. This table reports the results from the following time-series regression:

$$\beta_{L,large,t} - \beta_{L,small,t} = \alpha + b_1 t + b_2 (Own_{large,\,t-1} - Own_{small,\,t-1}) + b_3 (Size_{large,t-1} - Size_{small,\,t-1}) + b_4 (Illiq_{large,t-1} - Illiq_{small,\,t-1})$$

where $\beta_{L,large,t}$ and $\beta_{L,small,t}$ are the averages of quarterly market liquidity beta for large and small firms, respectively, and $Own_{large}$ and $Own_{small}$ are the averages of the number of shares held by the investor type (institutions, foreign institutions or local institutions) divided by number of shares outstanding across large and small firms, computed at the end of the previous quarter. $Size_{large,t-1}$ and $Size_{small,\,t-1}$ are the averages of the natural logarithm of the firm's market capitalization (in millions) across large and small firms, computed at the end of the previous quarter. $Illiq_{large,t-1} - Illiq_{small,\,t-1}$ are the averages of daily Amihud (2002) illiquidity measure across large and small firms, estimated over the previous quarter. The table presents the coefficient estimates. Numbers in brackets are $t$-statistics, adjusted for heteroskedasticity and autocorrelation using Newey and West (1987) standard errors with two lags.

| Variables | [1] | [2] | [3] | [4] | [5] | [6] | [7] |
|---|---|---|---|---|---|---|---|
| Intercept | 0.288 | 1.256 | 0.256 | 0.659 | 0.555 | 1.133 | 1.165 |
| | [1.80] | [2.14] | [1.44] | [1.14] | [3.60] | [1.20] | [1.44] |
| Institutional Ownership Spread | 1.375 | 1.945 | | | | | |
| | [2.42] | [3.16] | | | | | |
| Foreign Institutional Ownership Spread | | | 1.639 | 1.571 | | | 1.965 |
| | | | [2.92] | [2.96] | | | [3.04] |
| Local Institutional Ownership Spread | | | | | -0.596 | 0.161 | 1.704 |
| | | | | | [-0.60] | [0.11] | [1.26] |
| Size Spread | | -0.210 | | -0.076 | | -0.114 | -0.193 |
| | | [-1.78] | | [-0.75] | | [-0.63] | [-1.23] |
| Illiq Spread | | 0.001 | | 0.001 | | 0.002 | 0.001 |
| | | [0.24] | | [0.09] | | [0.28] | [0.21] |

**Table 7**

High ownership liquidity beta of quintile portfolios sorted by foreign institutional ownership. Panel A reports the average $\beta_{HI}$ sorted by foreign institutional ownership. Each quarter we sort stocks into quintiles based on the number of shares held by the foreign institutions divided by the total number of shares outstanding at the end of the previous quarter. For each quintile we report average $\beta_{HI}$ and the difference in mean $\beta_{HI}$ between the top and the bottom quintiles. $\beta_{HI}$ is estimated for foreign institutional investors as in Eq. 7. Panel B presents the dependent sort results. We first sort stocks into terciles based on the market capitalization at the end of the previous quarter: small (bottom tercile), mid-cap (middle tercile) or large (top tercile). Within each size group, we sort stocks into quintiles based on the foreign institutional ownership. Numbers in brackets are *t*-statistics, adjusted for heteroskedasticity and autocorrelation using Newey and West (1987) standard errors with two lags.

Panel A: One-way sort by foreign institutional ownership

| Foreign Institutional Ownership | | | | | | |
|---|---|---|---|---|---|---|
| Lo | 2 | 3 | 4 | Hi | Hi - Lo | *t*-stat |
| 0.025 | 0.02 | 0.081 | 0.105 | 0.142 | 0.117 | [5.37] |

Panel B: Two-way sort by size and foreign institutional ownership

| | Foreign Institutional Ownership | | | | | | |
|---|---|---|---|---|---|---|---|
| | Lo | 2 | 3 | 4 | Hi | Hi - Lo | *t*-stat |
| Large | 0.103 | 0.151 | 0.16 | 0.165 | 0.163 | 0.06 | [2.02] |
| Mid-Cap | 0.051 | 0.019 | 0.043 | 0.058 | 0.107 | 0.055 | [1.52] |
| Small | -0.001 | -0.008 | -0.019 | 0.008 | 0.029 | 0.03 | [1.15] |

**Table 8**
High ownership liquidity beta and foreign institutional ownership in the cross-section. This table reports estimates from Fama and Macbeth (1973) specification of the following regression:

$$\beta_{HI,i,t} = \alpha + b_1 Fown_{i,t-1} + b_2 Size_{i,t-1} + b_3 Illiq_{i,t-1} + \upsilon_{i,t}$$

where $\beta_{HI,i,t}$ is estimated for foreign institutional investors as in Eq. 7, $Fown_{i,t-1}$ is the foreign institutional ownership and computed as the number of shares held by foreign institutional investors divided by the total shares outstanding, $Size_{i,t-1}$ is the natural logarithm of the firm's market capitalization (in millions) and $Illiq_{i,t-1}$ is the daily average of the firm's Amihud (2002) illiquidity measure. $Fown_{i,t-1}$, $Size_{i,t-1}$, $Illiq_{i,t-1}$ are computed at the end of the previous quarter. We sort firms into terciles each quarter based on the market capitalization at the end of the previous quarter: small (bottom tercile), mid-cap (middle tercile) or large (top tercile). We present the time-series averages of the coefficients for all stocks as well as for each size group. Numbers in brackets are *t*-statistics, adjusted for heteroskedasticity and autocorrelation using Newey and West (1987) standard errors with two lags.

| | All Stocks | | Large | | Mid-Cap | | Small | |
|---|---|---|---|---|---|---|---|---|
| Variables | [1] | [2] | [3] | [4] | [5] | [6] | [7] | [8] |
| Intercept | 0.001 | -0.117 | 0.094 | 0.112 | 0.012 | -0.192 | -0.01 | -0.068 |
| | [0.05] | [-4.85] | [2.82] | [1.49] | [0.62] | [-1.32] | [-0.41] | [-0.76] |
| Foreign Institutional Ownership | 0.233 | 0.113 | 0.123 | 0.078 | 0.141 | 0.102 | 0.067 | 0.063 |
| | [6.40] | [3.20] | [2.70] | [2.04] | [2.07] | [1.54] | [0.97] | [1.02] |
| Size | | 0.026 | | 0.001 | | 0.041 | | 0.015 |
| | | [7.25] | | [0.14] | | [1.45] | | [0.77] |
| Illiquidity | | -0.001 | | -0.424 | | -0.015 | | -0.002 |
| | | [-0.50] | | [-1.51] | | [-0.84] | | [-0.30] |

**Table 9**

Correlated trading: ASX200 vs. non-ASX200 stocks. This table reports summary statistics on a number of proxies for correlated trading for ASX200 stocks and non-ASX200 stocks. We split the non-ASX200 stocks into medium and small groups based on their market capitalizations at the end of the previous quarter. Medium group represents top 50% of the non-ASX200 stocks and small group represents bottom 50% of the non-ASX200 stocks. $\beta_L$ is the market liquidity beta estimated as in Eq. 2. $\beta_{TO}$ is the trading volume comovements measure or volume beta and computed as follows. Each quarter, we use a market model for volume and regress the percentage change in the daily turnover ratio for the individual stock on the concurrent percentage change in the market wide, value-weighted turnover. Dollar Volume and Turnover are the average daily dollar volume and turnover, respectively, and Autocorrelation is the daily return autocorrelations. The columns present pooled mean, median, minimum, 25th percentile, 75th percentile, maximum and standard deviation of the variables for all firms.

| | | Variables | Mean | Median | Min | P25 | P75 | Max | Std Dev |
|---|---|---|---|---|---|---|---|---|---|
| ASX 200 Stocks | | $\beta_L$ | 0.412 | 0.413 | -5.364 | 0.059 | 0.757 | 3.532 | 0.568 |
| | | $\beta_{TO}$ | 0.795 | 0.814 | -1.541 | 0.54 | 1.056 | 3.093 | 0.406 |
| | | Dollar Volume ($thousands) | 21273.06 | 6174.85 | 27.234 | 2259.25 | 18222.26 | 781128.97 | 47832.52 |
| | | Turnover (%) | 0.397 | 0.341 | 0.015 | 0.233 | 0.493 | 5.537 | 0.267 |
| | | Autocorrelation (%) | -3.17 | -3.084 | -61.705 | -12.879 | 6.805 | 58.481 | 14.72 |
| Non-ASX 200 | All | $\beta_L$ | 0.085 | 0.088 | -5.782 | -0.364 | 0.541 | 5.445 | 0.77 |
| | | $\beta_{TO}$ | 0.293 | 0.308 | -3.792 | -0.13 | 0.73 | 4.117 | 0.701 |
| | | Dollar Volume ($thousands) | 602.764 | 166.722 | 3.321 | 68.377 | 442.052 | 101338.31 | 2465.24 |
| | | Turnover (%) | 0.255 | 0.161 | 0 | 0.084 | 0.299 | 7.98 | 0.358 |
| | | Autocorrelation (%) | -8.388 | -8.222 | -76.419 | -19.637 | 2.878 | 94.05 | 16.56 |
| | Medium | $\beta_L$ | 0.119 | 0.12 | -4.885 | -0.304 | 0.546 | 5.445 | 0.727 |
| | | $\beta_{TO}$ | 0.377 | 0.385 | -3.684 | -0.024 | 0.78 | 4.117 | 0.67 |
| | | Dollar Volume ($thousands) | 1050.4 | 370.413 | 3.691 | 171.13 | 818.668 | 101338.31 | 3409.16 |
| | | Turnover (%) | 0.197 | 0.134 | 0 | 0.066 | 0.254 | 5.237 | 0.225 |
| | | Autocorrelation (%) | -6.413 | -6.139 | -76.386 | -17.575 | 4.908 | 94.05 | 16.559 |
| | Small | $\beta_L$ | 0.051 | 0.051 | -5.782 | -0.421 | 0.535 | 4.722 | 0.809 |
| | | $\beta_{TO}$ | 0.21 | 0.213 | -3.792 | -0.239 | 0.669 | 3.75 | 0.721 |
| | | *Dollar Volume* ($thousands) | 154.116 | 80.78 | 3.321 | 42.027 | 161.253 | 21241.92 | 340.102 |
| | | *Turnover (%)* | 0.313 | 0.189 | 0 | 0.106 | 0.353 | 7.98 | 0.447 |
| | | *Autocorrelation (%)* | -10.368 | -10.418 | -76.419 | -21.47 | 0.454 | 64.101 | 16.323 |